\documentclass[a4paper]{article}
\usepackage{authblk}

\usepackage{times}
\usepackage{amsmath}    
\usepackage{amsfonts,bbm,bm}   
\usepackage{amssymb}
\usepackage{graphicx}
\usepackage{overpic}
\usepackage{caption,subcaption}
\usepackage{longtable}
\usepackage{setspace}
\usepackage[utf8]{inputenc}
\usepackage{xspace}
\usepackage{diagbox}

\usepackage{microtype}     
\usepackage[colorlinks = true,
            linkcolor = blue,
            urlcolor  = blue,
            citecolor = blue,
            anchorcolor = blue]{hyperref}   

\usepackage{multirow}
\usepackage[table,xcdraw]{xcolor}
\usepackage{eurosym}
\usepackage[final]{changes}

\usepackage{fancyref}
\usepackage{xcolor}
\usepackage{subcaption}
\usepackage{booktabs}
\usepackage{xltabular}
\usepackage{mathtools}
\usepackage{gensymb} 

\newcommand{\No}{{\mathcal{N}}}

\newcommand{\Fk}{\hat{F}_{{\mathrm K}}}
\newcommand{\Fks}{\hat{F}_{{\mathrm K};s}}
\newcommand{\Fkst}{\hat{F}_{{\mathrm K};s,t}}

\newcommand{\fk}{\hat{f}_{{\mathrm K}}}

\DeclareMathOperator{\erf}{erf}
\DeclareMathOperator{\sgn}{sgn}

\newcommand{\ind}{\mathbb{I}}
\newcommand{\om}{\omega}

\newcommand{\delt}{\delta t}

\newcommand{\ui}{\tilde{u}_{i}}

\providecommand{\keywords}[1]
{
  \small	
  \textbf{\textit{Keywords---}} #1
}
\begin{document}

\title{Non-parametric Kernel-Based Estimation and Simulation of Precipitation Amount}

\author[1]{Andrew Pavlides\thanks{\texttt{apavlidis@isc.tuc.gr}}}
\author[1]{Vasiliki D. Agou\thanks{\texttt{vagou@isc.tuc.gr}}}
\author[2]{Dionissios T. Hristopulos\thanks{\texttt{dchristopoulos@ece.tuc.gr};  Corresponding author}\hspace{2pt}}

\affil[1]{School of Mineral Resources  Engineering,  Technical University of Crete, Chania, Crete 73100, Greece}
\affil[2]{School of Electrical and Computer  Engineering,  Technical University of Crete, Chania, Crete 73100, Greece}

\maketitle

\date{}

\begin{abstract}
The probability distribution of precipitation amount strongly depends on  geography, climate zone,  and time scale considered. Closed-form parametric probability distributions are not sufficiently flexible  to provide accurate and universal models for precipitation amount over different time scales. In this paper we derive  non-parametric estimates of the cumulative distribution function (CDF)  of precipitation amount for wet periods.  The CDF estimates are obtained by integrating the kernel  density  estimator leading to semi-explicit CDF  expressions for different kernel functions. We investigate  an adaptive plug-in bandwidth (KCDE), using both synthetic data sets and reanalysis  precipitation data from the Mediterranean island of Crete (Greece). We show that KCDE  provides better estimates of the probability distribution  than the standard empirical (staircase)  estimate and kernel-based estimates that use the normal reference bandwidth.  We also demonstrate that KCDE enables the simulation of non-parametric precipitation amount distributions by means of the inverse transform sampling method.
\end{abstract}

\keywords{precipitation, kernel  estimation, simulation, non-Gaussian, reanalysis data, non-parametric estimate}

\newpage

\section{Introduction}

Precipitation is a key component of the hydrologic cycle.  It is particularly important for the health of ecosystems and the well-being of human societies, since it helps to recharge aquifers and  surface water bodies thus contributing to economic activity and the preservation of biodiversity.
On the other hand, modeling precipitation is not an easy task.  Features such as the intensity and  duration of precipitation vary significantly depending on geographic location, climate zone, season, and local topography.  There are also strong interannual variations in the amount of precipitation over any given geographical area.

Since precipitation is an inherently stochastic process, its properties are modeled by means of probability distributions.  Reflecting the highly variable features of precipitation, such probability distributions vary significantly depending on the location and the time scales of the measurements, as well as the modalities by means of which the data are obtained (e.g., rain gauges, satellite sensors, radar-based estimates, reanalysis methods).

To further complicate the situation, climate change has an impact on the precipitation patterns globally. It affects the duration and rate of precipitation events, the return periods, and the total amount of precipitation received by an area.  This is a major cause of concern, since altered precipitation patterns can destabilize ecosystems, cause significant economic damage due to extreme events, and ruin local  economies.  The risk is  greater in certain vulnerable areas of the globe which include the Mediterranean basin.  Precipitation plays a crucial role in the life of the major  Mediterranean islands (e.g., Sicily, Sardinia,  Cyprus, Corsica, and Crete). Changes in the intensity and total amount of precipitation, in combination with increased demand for water due to tourism and agricultural activities, can lead to soil erosion, depletion of groundwater aquifers, extinction of flora and fauna, increased flooding, salt water intrusion into aquifers,   and eventually desertification. Developing accurate, quantitative, data-based models for precipitation is crucial for predicting   climate change impacts and proposing effective policies  to mitigate them.

The intensity (rain rate) and total amount of precipitation over a specified aggregation time window are often modeled using parametric approaches. Time windows of one hour, one day during the wet period, an entire month, an entire season, or one year are used. Common parametric models  include the exponential, mixed exponential, gamma, lognormal, generalized gamma,  Weibull, generalized extreme value (GEV), as well as hybrid  mixtures of exponential with a Pareto tail~\cite{Li13,Ye18}.

The gamma distribution is a popular choice, regardless of the observation time window over which data were aggregated~\cite{Ye18}. It has been used to model daily rain rates~\cite{Choetal_2004}, as well as monthly and seasonal precipitation anomalies~\cite{Wilks90,wilks_1992}. It adequately models the variation of precipitation amounts in  wet periods with a scale parameter proportional to the number of  days and a shape parameter independent of the length of the observation window~\cite{Ison71}.  Zero precipitation values are taken into account in the gamma model using Type I censoring of the distribution on the left~\cite{Wilks90}. Left censoring implies that the number of values below a given threshold is considered known, but the exact values are not. The same approach can also be used for the other distributions discussed below.
However, recent studies have highlighted that the tails of the gamma distribution are not sufficiently heavy for heave rain events~\cite{Neratzaki19,Toumi05}. This is due to the fact that the exponential part of the gamma tail dominates the power-law term, and thus in many cases the gamma tail behaves similarly to the exponential.

The lognormal distribution has been used to approximate rain-rate  \cite{Biondini1979,Kedem1987,Sauvageot1994},  cumulus cloud populations~\cite{Lopez1977},
hourly precipitation~\cite{Shoji06}, storm height distribution over the tropical oceans~\cite{Short00},  amount of  precipitable water~\cite{Foster03,Fosteretal2006}, and the estimated  average rain rate  from satellite observations~\cite{Kedem90}.
A recent study of the lognormal and gamma distributions  concluded that both  fit satisfactorily  TRMM data (obtained from the Tropical Rainfall Measuring Mission research satellite which operated from 1997 to 2015) of  daily average rain-rate~\cite{Choetal_2004}.

The Generalized Extreme Value (GEV) distribution has  been  applied to modeling of precipitation amount and extremes over different time scales, such as maximum one-day rainfall~\cite{Wangetal2017,Coles2001}, $k$-day extreme precipitation (for $k$ ranging between 1 and 30)~\cite{Gellens02},  daily precipitation forecasts~\cite{Scheuerer14}, annual maximum precipitation~\cite{Koutsoyiannis2004}, monthly return levels and their annual cycle~\cite{Schindleretal2012}, and  seasonal  extremes using ensembles of climate models~\cite{Kharinetal2007}.
Distribution models  that allow more flexible tail behavior than the GEV are considered in~\cite{PAPALEXIOU2012AWR,RHO2019AWR,Papalexiou20}.

Parametric models offer the advantage of closed-form expressions for the probability density function, and in most cases for all the main probability functions (i.e., cumulative distribution, survival, and hazard rate functions). In addition, they are motivated by theoretical arguments. For example,  the Fisher–Tippett–Gnedenko theorem  states that the maximum of a sample of independent, identically distributed random variables (after suitable affine transformation)  converges in distribution to the GEV family  (which includes the Gumbel,  Fréchet, and reverse Weibull models)~\cite{Fisher28,Gnedenko43}.
A recent analysis based on  stochastic modeling of the competing factors in the atmospheric moisture budget justifies the use of the gamma distribution  for precipitation intensity~\cite{MV19}.

In spite of their considerable advantages, parametric models have a number of shortcomings.  One of them is their limited flexibility.  For example, a recent study of daily precipitation values over Europe showed that the gamma distribution is not supported by rigorous statistical testing at more than 40\% of the examined stations~\cite{Vlcek09}.  In certain cases, a parametric expression of the cumulative distribution function (CDF)  cannot be easily deduced. Such an example is the amount of monthly precipitation in semi-arid Mediterranean areas~\cite{AgouMTh}.

Standard parametric probability models (e.g., exponential, gamma distribution) may also fail to capture both the low-value (left tail) and high-value (right tail) regimes of the data  distribution. This has motivated the introduction of hybrid models, which combine more than one distribution types.
A different approach   considers flexible three-parameter distributions that allow independent control of both tails~\cite{Papalexiou20}.
Another problem is that parametric models often assume stationarity, i.e., that the parameters of the distribution remain constant during the time of observation. However, in light of the accelerating effects of climate change on meteorological variables such as temperature and precipitation~\cite{wmo21},  the stationarity assumption is not easily justified~\cite{Madsen14}. Moreover, parametric models that have  proved satisfactory in the past may not adequately capture precipitation patterns in future climate. This concern is fueled by the expected increase of extreme precipitation events~\cite{Madsen14}, as well as climate model simulations predicting changes in the  statistical patterns of daily precipitation extremes~\cite{MV21}.

Non-parametric approaches that estimate the probability distribution from the data without reference to a specific mathematical form provide a useful alternative to parametric models. The simplest non-parametric estimator is the empirical CDF, which has the form of a non-decreasing step function (a staircase). However, the empirical CDF is discontinuous. This is unsatisfactory  for continuously-valued variables such as the rain rate, especially for  limited sample size.  In addition, the staircase estimate is not suitable for the simulation of continuous variables.

Kernel density (KDE) estimation (known as the  Parzen Rosenblatt window method) is a non-parametric method which constructs approximations of the probability density function (PDF) without step discontinuities~\cite{Rosenblatt56,Parzen62}.
Kernel density estimators employ a bandwidth parameter $h$ which determines the range of the kernel function and controls the smoothness of the estimated PDF~\cite{Ghosh18}.
Selection of an optimal KDE bandwidth  is an important problem  in statistics ~\cite{Park90,Wand94,Jones96}. Plug-in bandwidths  rely on {\it a priori} assumptions about the probability distribution of the data; they are derived by  minimizing the Asymptotic Mean Integrated Square Error  between the KDE and the unknown PDF~\cite{Sheather91,Wand94,Jones96,Ghosh18}. For example, if the data  follow the normal probability distribution $\No(m,\sigma^2)$, the \emph{normal reference rule (rule of thumb)}   bandwidth $h$ for the Gaussian kernel is  $h=1.06 \sigma N^{-1/5}$, where $N$ is the data size~\cite{scott2015}.

Non-parametric approaches have been used in precipitation studies. For instance, \cite{SHARMA1999,Harrold2003WR} reported on KDE and its application in the  simulation of daily precipitation. The study~\cite{Mosthaf17} reports on non-parametric precipitation distribution modeling using the Gaussian kernel and numerical integration of the density.
The normal reference rule bandwidth with the interquartile range as measure of scale~\cite{scott2015}
and  the Sheather-Jones plug-in bandwidth~\cite{Sheather91} were tested and returned comparable results.

We propose a non-parametric, kernel-based estimator (KCDE) which targets the CDF instead of the PDF. We analytically evaluate the kernel integrals which form the KCDE building blocks  for selected kernel functions with both compact and infinite support. Based on synthetic data sets, we show that  KCDE equipped with the plug-in BGK bandwidth derived by Botev, Grotowski, and Kroese~\cite{Botev10} estimates the data probability distribution more accurately than the normal reference bandwidth which is typically used in KDE.  In addition,  KCDE allows the efficient simulation of precipitation amounts from the non-parametric CDF by means of the inverse transform sampling method.  We test KCDE performance    against the empirical CDFs for several kernel functions.

Our analysis employs synthetic precipitation data motivated by rain gauge measurements as well as ERA5 reanalysis products for the Mediterranean island of Crete (Greece). Crete is  a major Mediterranean island (fifth in size overall and the largest island of Greece), where the combination of adverse climate change forecasts, and  increasing tourist and agricultural activities poses serious threats to the sustainability of the island's water resources~\cite{Varouchakis18,Agou2019,Nerantzaki20}.

The remainder of this paper is structured as follows: Section~\ref{sec:methods} describes the methods used and a  short synopsis of the investigated probability distribution models.
Section~\ref{sec:theory} presents the theoretical developments proposed in this work. The KCDE is derived and the integrals involved in kernel-based CDF estimates, $\hat{F}_k(z)$, are evaluated for selected kernel functions in Sections~\ref{ssec:CDF-estimation}-\ref{ssec:GaussIntegral} (with  more details  in~\ref{appA}). The BGK plug-in bandwidth estimation  by~\cite{Botev10}  is reviewed in Section~\ref{ssec:bgk-band} (algorithmic details are given in~\ref{appB}).
Finally, a simulation method for precipitation amounts based on KCDE is proposed in Section~\ref{ssec:simulations}. Section~\ref{sec:results} presents applications of KCDE (using the BGK bandwidth) to synthetic data  (Section~\ref{sec:synth_data} and ~\ref{appC}) and to ERA5 reanalysis data (Section~\ref{ssec:era5} and~\ref{appD}).
Discussion, conclusions and suggestions for further studies are  given in Section~\ref{sec:conclusions}.

\section{Methodology}
\label{sec:methods}

\subsection{Stochastic modeling of precipitation amount}
\label{ssec:precipitation-stochastic}

In the following, we assume that the precipitation rate is a continuous-time, stationary stochastic process $R(t;\om)$ defined over a probability space $(\Omega, \mathcal{F}, P)$, where $\om \in \Omega$ is the state index in the sample space $\Omega$, $\mathcal{F}$ is a $\sigma-$algebra of events, and $P$ is the probability measure for these events~\cite{Papoulis02,dth20}.

The data comprise precipitation amounts $\left( z_{1}, z_{2}, \ldots, z_{N}\right)$ observed at  times $t_{1}, t_{2}, \ldots, t_{N}$.  The times $t_{n}= n \delt$ are multiples of a characteristic time step $\delt$,  corresponding to one day, week, month, season, or year. The amounts $z_{n}$  represent the cumulative amount of precipitation over the time interval $[t_{n-1}, t_{n})$. We focus on wet time intervals, i.e.,  $z_{n} > 0$ for all $n$. The cumulative precipitation represents samples of the following stochastic integral (for definition see~\cite{Papoulis02}):
\[
Z_{\delt}(t_{n};\om)= \int_{t_{n-1}}^{t_{n}} dt\, R(t;\om), \; n=1, \ldots, N.
\]

In the following, for brevity we refer to $Z_{\delt}(t_{n};\om)$ as $Z(t;\om)$. However, $Z(t;\om)$  implicitly depends on the averaging window $\delt$.
Multiple samples  from the same probability model will be distinguished by means of the index $m=1, \ldots, M$, i.e.,
$Z^{(m)}_{N}=\left( z_{1}^{(m)}, z_{2}^{(m)}, \ldots, z_{N}^{(m)}\right)^\top$, where  $\top$ denotes the transpose of a matrix or vector.

Different probability models will be distinguished by means of the integer subscript $j$, i.e., $Z^{(m)}_{N;j}$, $j=1, \ldots, J$ and $J$ is the number of models tested.
The marginal CDF of $Z(t;\om)$ is defined by $F(z)= {\mathrm{Prob}}(Z \le z)$ and the marginal PDF by $f(z)=\dfrac{dF(z)}{dz}$.

\subsection{Probability distribution estimation}
The respective estimates of the CDF and PDF from the data are denoted by  $\Fk(z)$,  $\fk(z)$ if the estimates are based on KDE and $\hat{F_s}(z)$,   $\hat{f_s}(z)$ if the estimates are obtained by means of the empirical staircase function.

A  smoothing kernel  is a real-valued, non-negative function $K(z-z';h)= K(\frac{z-z'}{h})$,  which respects the properties of symmetry, i.e., $K(u)=K(-u)$ and normalization, i.e.,
$\int_{-\infty}^{\infty} du \, K(u)=1$.  The parameter $h>0$ is the kernel bandwidth. In addition,  the maximum of $K(u)$ is obtained for $u=0$ and $K(u)$ decreases monotonically with $\lvert u \rvert$. This behavior ensures that when estimating the PDF at $z$, the kernel assigns higher weight to a data value $z_n$ than to a value $z_{m}$,  if $z_n$ is closer to the target value $z$ than to $z_m$.

\subsection{Probability distribution models for precipitation data}
\label{ssec:Distributions}

We investigate four non-Gaussian probability distributions commonly associated with precipitation modeling~\cite{Fosteretal2006}: the gamma (GAM), the  Generalized Extreme Value (GEV), the lognormal (LGN), and the Weibull (WBL) models. The PDFs of these distributions are given in Eq.~\eqref{eq:model-pdf} below.

\begin{subequations}
\label{eq:model-pdf}
\begin{equation}
\label{eq:gamma-pdf}
\mbox{gamma:} \;  f(z;\xi,\sigma)=  \dfrac{z^{\xi-1}e^{\tiny {-\frac{z}{\sigma}}}}{\sigma^{\xi}\Gamma(\xi)} , \quad  \ z>0  \, \textrm{ and } \ \xi,\sigma>0,
\end{equation}
\begin{align}
\label{eq:gev-pdf}
\mbox{GEV:} \; & f(z;\mu,\sigma,\xi)=  \dfrac{1}{\sigma}\, y(z) ^{\xi+1} \exp\left[-y(z)\right], \quad z \in \mathbb{S}(\xi),
\nonumber
\\[1ex]
& \textrm{where } y(z) =  \begin{cases} \left[ 1+\xi\left(\frac{z-\mu}{\sigma}\right)\right] ^{-1/\xi} & \xi\neq 0,  \\[1ex]
\exp \left(-\frac{z-\mu}{\sigma}\right) & \xi=0, \end{cases}
\end{align}
\begin{equation}
\label{eq:logno-pdf}
\mbox{lognormal:} \;  f(z;\mu,\sigma) = \frac{1}{z \sigma \sqrt{2\pi}} \, \exp \left\{ {\frac{-(\log \, z - \mu)^2}{2\sigma^2}} \right\} ,  \quad  z>0, \; \sigma>0.
\end{equation}

\begin{equation}
    \mbox{Weibull:} \;  f(z;\sigma,\xi) =
\begin{cases}
\frac{\xi}{\sigma}\left(\frac{z}{\sigma}\right)^{\xi-1}e^{-(z/\sigma)^{\xi}}, & z\geq0 ,\\
0, & z<0,
\end{cases}
\label{eq:weibull-pdf}
\end{equation}
\end{subequations}

\medskip

\noindent In the above equations, $\mu\in\mathbb R$ is the location parameter, $\sigma > 0$ is the scale parameter that controls the dispersion of the distribution, and $\xi$ is a shape parameter.  $\Gamma(\cdot)$ is  the gamma function.
Note that for $\xi=1$, the gamma distribution defaults to the exponential model.

In the case of the GEV distribution,  $\xi$ determines which type (I, II, or III)  is represented.  The Gumbel distribution (type I) is obtained for $\xi=0$, the Fr\'{e}chet distribution (type II) for $\xi<0$, and the  Reverse-Weibull distribution (type III) for $\xi>0$~\cite{Haan.Ferreira2010}. For  $\xi \in (-0.278, 1)$ the GEV distribution is positively skewed;  it also has a finite mean given by $m=\mu + \sigma \frac{\Gamma(1-\xi)-1}{\xi}$ for $\xi \neq 0$, and $m=\mu + \sigma\gamma_{E}$ for $\xi=0$, where $\gamma_{E} \approx 0.5772$ is the Euler-Mascheroni constant~\cite{Scheuerer14}.
The support $\mathbb{S}(\xi)$  of the GEV distribution is $\mathbb{S}(\xi)=[\mu - \sigma/\xi, \infty)$ for $\xi>0$, $\mathbb{S}(\xi)=( - \infty, \infty)$ for $\xi=0$, and $\mathbb{S}(\xi)=(-\infty, \mu - \sigma/\xi ]$ for $\xi<0$.

The respective CDFs of the gamma, GEV, lognormal and weibull models are given below:
\begin{subequations}
\label{eq:model-cdf}
\begin{equation}
\label{eq:gamma-cdf}
\mbox{gamma:} \,  F(z;\xi,s)= \int\limits_0^z  f(u;\xi,s) \: \mathrm{d}u=\dfrac{{\gamma}(\xi,\frac{z}{s})}{\Gamma(\xi)}, \;
\end{equation}

\begin{equation}
\label{eq:gev-cdf}
\mbox{GEV:} \;  F(z;\mu,\sigma,\xi)= \exp \left[- y(z)\right], \quad z \in \mathbb{S}(\xi),
\end{equation}

\begin{equation}
\label{eq:logno-cdf}
\mbox{lognormal:} \,  F(z;\mu,\sigma) = \frac{1}{2} \left\{ 1 + \erf \left( \frac{\log \, z - \mu}{\sigma \sqrt{2}} \right) \right\},  \quad \textrm{for} \ z>0,
\end{equation}

\begin{equation}
    \mbox{Weibull:} \, F(z;\sigma,\xi) = 1 - e^{-(z/\sigma)^\xi}, \quad \textrm{for} \ z>0.
    \label{eq:weibul-cdf}
\end{equation}
\end{subequations}

\medskip

\noindent In Eq.~\eqref{eq:model-cdf}, the function $y(z)$ is defined in Eq.~\eqref{eq:gev-pdf},  ${\gamma}\left(\xi,\cdot\right)$ represents the lower incomplete gamma function, and $\erf(\cdot)$ is the error function~\cite{Abramowitz88}. Note that for the gamma and lognormal distributions $F(0;\cdot,\cdot)=0$, while for the GEV finite values of the CDF are possible for negative values of $z$. Since $z<0$ is not acceptable for precipitation amounts, typically one sets $F(z<0)=0$~\cite{Scheuerer14}.

\section{Theory and Calculations}
\label{sec:theory}
In this section we derive a non-parametric, kernel-based method for CDF estimation from the data, and we propose a simulation method for  precipitation amounts from the kernel-based CDF estimate.  Since we focus in cumulative  precipitation amounts over specific time windows, we only take account of non-zero precipitation values.

\subsection{Kernel-based CDF estimation}
\label{ssec:CDF-estimation}

Let $z_{[i]}$ represent the $i$-th value of the ordered sample, which satsfies the following properties: for any given $i \in \{1, \ldots,  N\}$ there exists $j \in \{1, \ldots,  N\}$   such that $z_{[i]}= z_{j}$ and
$z_{[1]} \le z_{[2]} \le \ldots \le z_{{[N]}}$.

The  standard PDF kernel density estimator  (KDE) is given by
\begin{equation}
\hat{f}_{\mathrm K}(z) = \dfrac{1}{Nh}\sum_{i=1}^{N}{K}\left(\dfrac{z-z_{[i]}}{h} \right),
\label{eq:kernel_PDF_estim}
\end{equation}
where $h$ is the kernel bandwidth of the kernel $K(\cdot)$.

Non-parametric estimates (KCDE) of the  CDF based on kernel functions $K(u)$ can also be obtained. These estimates are given by the following weighted sum

\begin{equation}
\Fk(z) = \sum\limits_{i=1}^{N}\dfrac{1}{N} \, \tilde{K}\left(\dfrac{z-z_{[i]}}{h} \right).
\label{eq:CDF_Kernel}
\end{equation}
In Eq.~\eqref{eq:CDF_Kernel},    $\tilde{K}\left(\cdot \right) $ is the \emph{CDF kernel step} defined by means of the  integral
\begin{equation}
\tilde{K}\left(\dfrac{z-z_{[i]}}{h} \right)  = \frac{1}{h} \int_{-\infty}^{z} dz' K\left(\dfrac{z'-z_{[i]}}{h} \right).
\label{eq:Kernel_integral}
\end{equation}
The CDF kernel steps are smoothed versions of the rectangular steps used in the empirical (staircase) CDF estimation.
Equation~\eqref{eq:CDF_Kernel} is obtained from Eq.~\eqref{eq:kernel_PDF_estim} using  the  integral $\Fk(z) = \int_{-\infty}^{z} \hat{f}_{\mathrm K}(z')\, dz'$.
Table~\ref{tab:Kernels} lists commonly used kernel functions and their respective  CDF  steps.  The CDF step of the Gaussian kernel is evaluated in Section~\ref{ssec:GaussIntegral}, while  the CDF steps of  other commonly used kernels  in~\ref{appA}.

A suitable kernel bandwidth $h$ can be estimated by  (i) minimizing specific cost functions such as the Root Mean Square Error (RMSE) or the Kullback–Leibler divergence (KLD) between $\Fk(\cdot)$ and $F(\cdot)$ using leave-one-out cross validation (LOOCV)  or (ii)  explicit (plug-in)  bandwidth formulas based on parametric assumptions~\cite{Chen17,Ghosh18}.  The use of plug-in bandwidths produces similar or better PDF estimates for non-uniformly distributed data  than bandwidths estimated with LOOCV~\cite{Mosthaf17, chu2015}.
The optimal bandwidth  in practice depends on the CDF shape and the data density (i.e., the spacing between  measured values)~\cite{benson2021}.
Our preliminary investigations which compared $\Fk(z)$ based on  (i) the BGK bandwidth and (ii) the LOOCV bandwidth for synthetic data exhibited superior performance of the BGK bandwidth.
Hence, we use the plug-in BGK bandwidth~\cite{Botev10} (see Section~\ref{ssec:bgk-band}).

In KDE, boundary values may require special treatment.
For example, large bandwidths can lead to  $\Fk(z)>0$ for $z<0$, which is not meaningful for precipitation amounts. Various solutions have been proposed for density estimation near the boundary~\cite{jones1993, malec2014}. Note that it is not permissible to  set the PDF equal to zero for $z<0$, because it would violate the normalization of the PDF.  Instead, we modify the CDF so that if $\Fk(0)=p >0$,  we set $\Fk(z)=0$ for $z<0$ and $\Fk(0)=p$.
This assignment implies zero probability for $z<0$ and a discontinuous jump at $z=0$, while it maintains normalization, i.e.,  $\Fk(z \to \infty)=1$.

\begin{table}[!ht]
\centering
\caption{Kernel functions $K(z)$ and respective CDF kernel steps, $\tilde{K}\left( u \right)$, as defined by means of Eq.~\eqref{eq:Kernel_integral}. $h$ is the kernel bandwidth.  $\ind(|z|\le 1)$ is an indicator function used for compactly supported kernels: $\ind(|z|\le 1)=1$ when $ \vert z \vert\leq1 $ and $\ind(|z|\le 1)=0$ otherwise.  $\sgn(u)=u/\lvert u \rvert$ is the sign function. The following kernel models are used:  Ga: Gaussian, Ex: Exponential, Ep: Epanechnikov, Bp: Bitriangular, Tr: Triweight, Sp: Spherical, Un: Uniform. For \emph{compact support kernels} (i.e., all but the exponential and Gaussian) the equations for $\tilde{K}(u)$  are valid for $ -1 \le u \le 1$; for $u<-1$ it holds that $\tilde{K}(u)=0$, while for $u>1$ one gets $\tilde{K}(u)=1$.}
\begin{tabular}{lll}
Kernel & $K(z)$   & $ \tilde{K}(u)$ \; (for compact kernels $ \lvert u \rvert \le 1 $)  \\[1ex]
\hline
\\
Ga  & $\exp(-z^2)$  &
$ \dfrac{1}{2} \left[ 1+ \erf(u) \right] $
\\[1ex]
Ex  & $\exp(-\vert z \vert)$  &
$   \sgn(u) \,
\dfrac{1 - e^{ - \vert u \vert }}{2 } +
\dfrac{1}{2}  $ \\[1ex]
Ep  & $ \dfrac{3}{4}(1-z^2)\, \ind(|z|\le 1) $  &
 $ \dfrac{3}{4}\,\left( u - \dfrac{u^3}{3} + \dfrac{2}{3} \right) $ \\[1ex]
Bp  & $ \dfrac{3}{2}(1-\vert z \vert)^2 \, \ind(|z|\le 1) $  &
$ \dfrac{3}{2} \,\left(   \dfrac{u^3}{3} - u^2\,\sgn(u)
+ u +\dfrac{1}{3} \right) $ \\[1ex]
Tr  & $(1-z^2)^3\, \ind(|z|\le 1)$  &
$   \dfrac{35}{32} \left( u -u^3 +\dfrac{3u^5}{5} -\dfrac{u^7}{7} +\dfrac{16}{35} \right)$  \\[1ex]
Sp  & $ \left(1 - 1.5 |z| + 0.5 |z|^3\right) \, \ind(|z|\le 1) $  &
 $ \dfrac{4}{3}  \left( \sgn(u)\,\dfrac{u^2-6}{8}\,u^2+u+ \dfrac{13}{8} \right) $ \\[1ex]
 Un  & $ \dfrac{z}{2}\, \ind(|z|\le 1) $  &
  $ \dfrac{1}{2}\, (u+1)  $ \\[1ex]
\hline
\end{tabular}
\label{tab:Kernels}
\end{table}

\subsection{CDF step for Gaussian kernel function}
\label{ssec:GaussIntegral}
The Gaussian  (squared exponential) kernel, defined by $K(u)=\frac{1}{\sqrt{\pi}}\exp(-u^2)$, is an infinitely extended function which has  wide-ranging applications in data analysis, machine learning, regression and other fields~\cite{xiao2014,kitayama2011,kusano2016}.

Using the change of variable
$u_{i}= (z-z_{[i]}) /h$
it follows that $dz =  h\, du_{i}.$
Subsequently, the CDF kernel step  $\tilde{K} \left(u \right)$ is evaluated by means of the integral

\begin{equation}
\label{eq:cdf-kernel-Gauss}
 \tilde{K} \left(u_{i} \right)  =   \int_{-\infty}^{u_{i}} dx~K(x) =  \dfrac{1}{2} \left[ \,\erf(u_{i}) +  1\right]\,,
 \end{equation}

\noindent  where $\erf(\cdot)$ is the Gauss error function
 \begin{equation*}
     \erf(u_{i}) = \dfrac{2}{\sqrt\pi}\int_0^{u_{i}} e^{-x^2}\,dx \,.
 \end{equation*}

\begin{figure}
\centering
\includegraphics[width=0.9\textwidth]{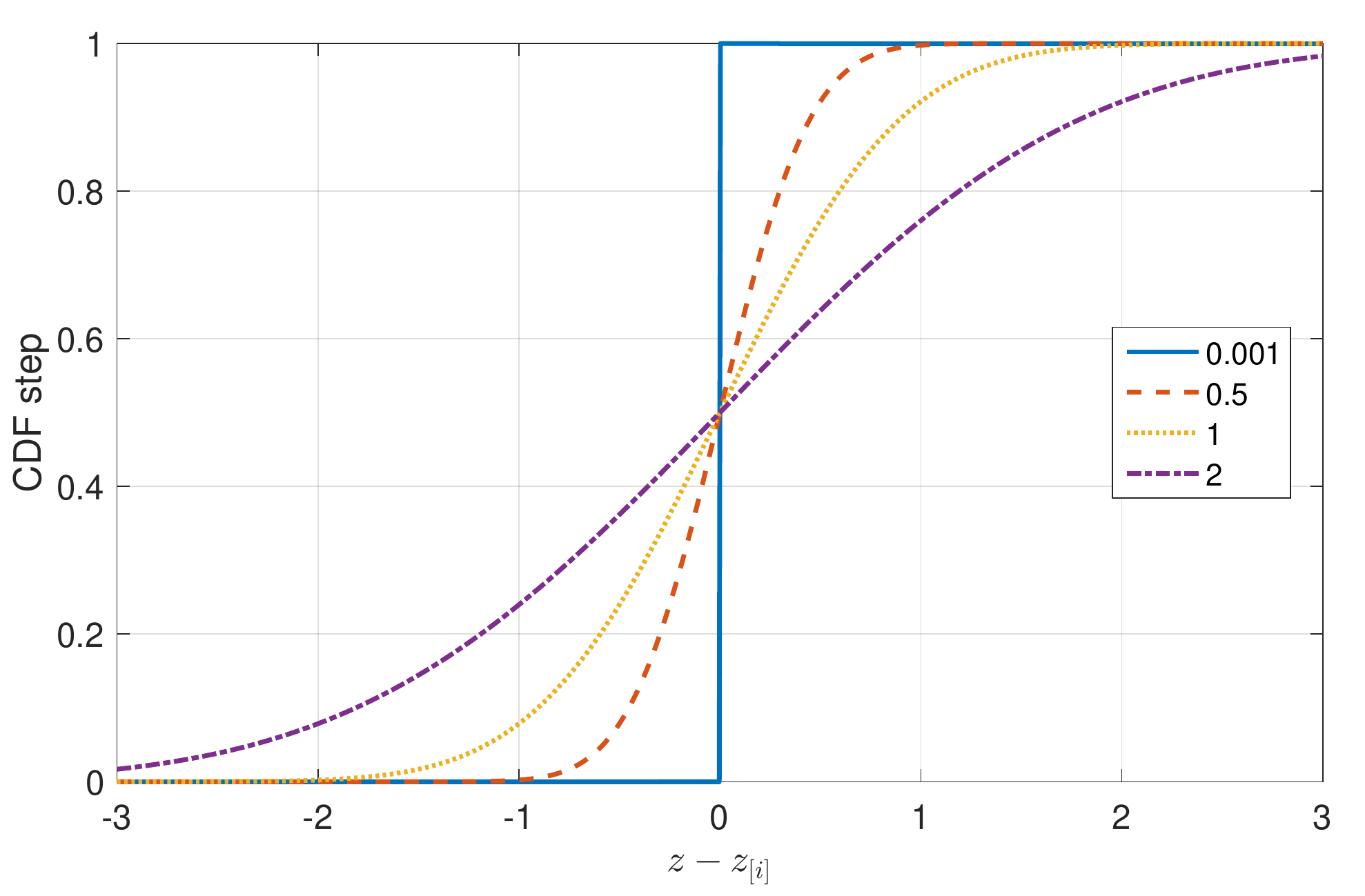}
\caption{CDF kernel step for the Gaussian kernel based on Eq.~\eqref{eq:cdf-kernel-Gauss} for different bandwidths.}
\label{fig:cdf_kernel_step}
\end{figure}

The CDF kernel step obtained from Eq.~\eqref{eq:cdf-kernel-Gauss}  is illustrated in Fig.~\ref{fig:cdf_kernel_step} for different $h$ values.  As shown in these plots, the CDF kernel step tends to the discontinuous step function as $h \to 0$; then KCDE tends to the empirical staircase estimator. As $h$ increases, the bandwidth becomes smoother.

Results for the CDF kernel steps corresponding to other kernel functions are presented in Table~\ref{tab:Kernels}; they are based on the calculations  in~\ref{appA}.
For compactly supported kernels, the step $\tilde{K}(u)$ is zero for $u<-1$, it is given by the expressions listed in Table~\ref{tab:Kernels} for $\lvert u \rvert \le 1$, and it is equal to one for $u>1$. For kernels with infinite support (e.g., Gaussian, exponential), the kernel step approaches zero and one asymptotically as $u \to -\infty$ and $u \to \infty$ respectively.

\subsection{BGK bandwidth estimation}
\label{ssec:bgk-band}

The standard  bandwidth selection in KDE   aims to minimize the Asymptotic Mean Integrated Square Error (AMISE)~\cite{Ghosh18, scott2015, benson2021}. Assume a Gaussian kernel $K(\cdot)$ and a PDF $f(z)$ which is at least twice differentiable, i.e.,  $f'(z)$ and $f''(z)$ are finite numbers for all $z$. Botev, Grotowski and Kroese~\cite{Botev10} (henceforward BGK), have shown that the  bandwidth which minimizes AMISE is  given by
\begin{equation}
\label{eq:h_opt}
h= \left(
\dfrac{1}{2 \sqrt{\pi} N \, G\left[f(\cdot) \right] }
\right)^{1/5}\, ,
\end{equation}
where $G\left[f(\cdot)\right]=\int_{-\infty}^{\infty}dz\,\left[ f''(z) \right]^2$ is a functional of $f(z)$.

Equation~\eqref{eq:h_opt}  in theory provides the BGK plug-in bandwidth. However, in practice it requires knowledge of the PDF $f(z)$,  which is in general unknown \emph{a priori}.  BGK approximates $f(z)$  by means of $\hat{f}_k(z)$, which is the KDE based on an auxiliary bandwidth $a$. The latter is calculated by means of an iterative algorithm after several repetitions~\cite{Botev10}. The details of the  algorithm are described in~\ref{appB}.
Note that in  BGK  different bandwidths are proposed for PDF versus CDF estimation. Our approach explicitly integrates the PDF kernel density estimates; so, it uses the PDF  bandwidth formula.

For data that deviate from the normal distribution, the normal reference rule can lead to  significant PDF oversmoothing ~\cite{Botev10, Ghosh18, Silverman1986, Sheather91}.
Another popular method of bandwidth estimation involves using LOOCV to minimise KLD or RMSE between $\Fk(\cdot)$ and $F(\cdot)$. LOOCV bandwidth estimates require time consuming numerical optimization and often yield small bandwidths that tend to produce peakedness (high kurtosis) in the density~\cite{Sheather2004, Mosthaf17}.

The BGK algorithm has two significant advantages over other plug-in methods: It does not require  normal data, and it is computed efficiently without the need for  numerical optimization~\cite{Botev10}.  Even though the BGK bandwidth is optimized for the Gaussian kernel, our tests with synthetic data in Section~\ref{sec:synth_data} show that it also performs well with compact-support kernels.

\subsection{KCDE-ITS simulation algorithm for precipitation amount}
\label{ssec:simulations}

Time series of precipitation amounts that respect KCDE-derived distributions  can be simulated using the \emph{inverse transform sampling (ITS) method}~\cite{dth20}.  Given a time series  of precipitation amounts $\left( z_{1}, z_{2}, \ldots, z_{N}\right)$, the goal is to generate an ensemble of time series, $\left( z_{1}^{(m)}, z_{2}^{(m)}, \ldots, z_{N'}^{(m)}\right)$, where $m=1, \ldots, M$ is the state index, that share the same probability distribution as the sample.  Henceforth we  drop the index $(m)$ to abbreviate notation.

The ITS method is based on the law of conservation of probability under nonlinear, monotonic transformations: If $Z(\om)$ is a random variable with CDF $\hat{F}_{Z}(z)$ and $U(\om)$ is a uniformly distributed random variable, then
\begin{equation}
\label{eq:rn-trans}
\hat{F}_{Z}(z)=F_{U}(u)=u \Rightarrow z = \hat{F}^{-1}_{Z}(u),
\end{equation}
where  $F_{U}(\cdot)$ is the CDF of the uniform distribution~\cite{Papoulis02,dth20}. The inverse CDF $\hat{F}^{-1}_{Z}(\cdot)$ can be calculated either by means of numerical optimization or a lookup table; we use the latter approach. For compactly supported kernel functions,
ITS simulation  involves the following steps:

\begin{enumerate}
\item The KCDE-based CDF of precipitation amount $\hat{F}_{Z}(z)$ is derived from the sample by applying Eq.~\eqref{eq:CDF_Kernel}.

\item A lookup table is generated which comprises a set of $L=10{,}000$ discretization points, i.e.,  $\{(z_l,u_l)\}_{l=1}^{L}$ pairs.

\item The set $\{z_{l}\}_{l=1}^L$ is obtained by uniformly distributing  $z_l \in [{z'}_{\min}, {z'}_{\max}]$, where ${z'}_{\min}=\max(eps,z_{\min}-h)$, ${z'}_{\max}=z_{\max}+h$,  $z_{\min}$, $z_{\max}$ are respectively the minimum and maximum values of the available sample, and $h$ is the BGK bandwidth.

\item The set $\{u_{l}\}_{l=1}^L$ is generated by setting $u_{l} = \hat{F}_{Z}(z_{l})$, for all $l=1, \ldots, L$ according to Eq.~\eqref{eq:rn-trans}.

\item A set of uniform random numbers $\{u_{n}\}_{n=1}^{N'}: \, u_{n} \overset{d}=U(0,1)$ are generated.

\item For each $u_n$, $n=1, \ldots, N'$
the nearest neighbor $u_{l \mid n}$ in the lookup table is found, and the simulated value $z_n$ is set equal to  $z_{l \mid n}$, i.e., which is the pair of $u_{l \mid n}$ in the lookup table.

\end{enumerate}

\section{Results}
\label{sec:results}
This section presents estimation of precipitation amount CDFs from synthetic and ERA5 reanalysis data using the proposed KCDE method. It also uses the KCDE-based simulation to generate and discuss probable scenarios of precipitation amounts.

\subsection{KCDE application to synthetic data}
\label{sec:synth_data}

We generate data from parametric  probability distributions with statistical properties  motivated by observations of monthly precipitation amounts on the Mediterranean island of Crete (Greece)~\cite{AgouMTh,Agou2019}.
One hundred  time series of independent values are generated from the gamma,  
GEV, Weibull and lognormal distributions discussed in Section~\ref{ssec:Distributions}. Samples of five different sizes ($N$=50, 100, 200, 500 and 1000 points) are generated  from each distribution.

\begin{figure} [h!]
  \centering
    \begin{subfigure}[h!]{0.49\textwidth}
        \includegraphics[width=\textwidth]{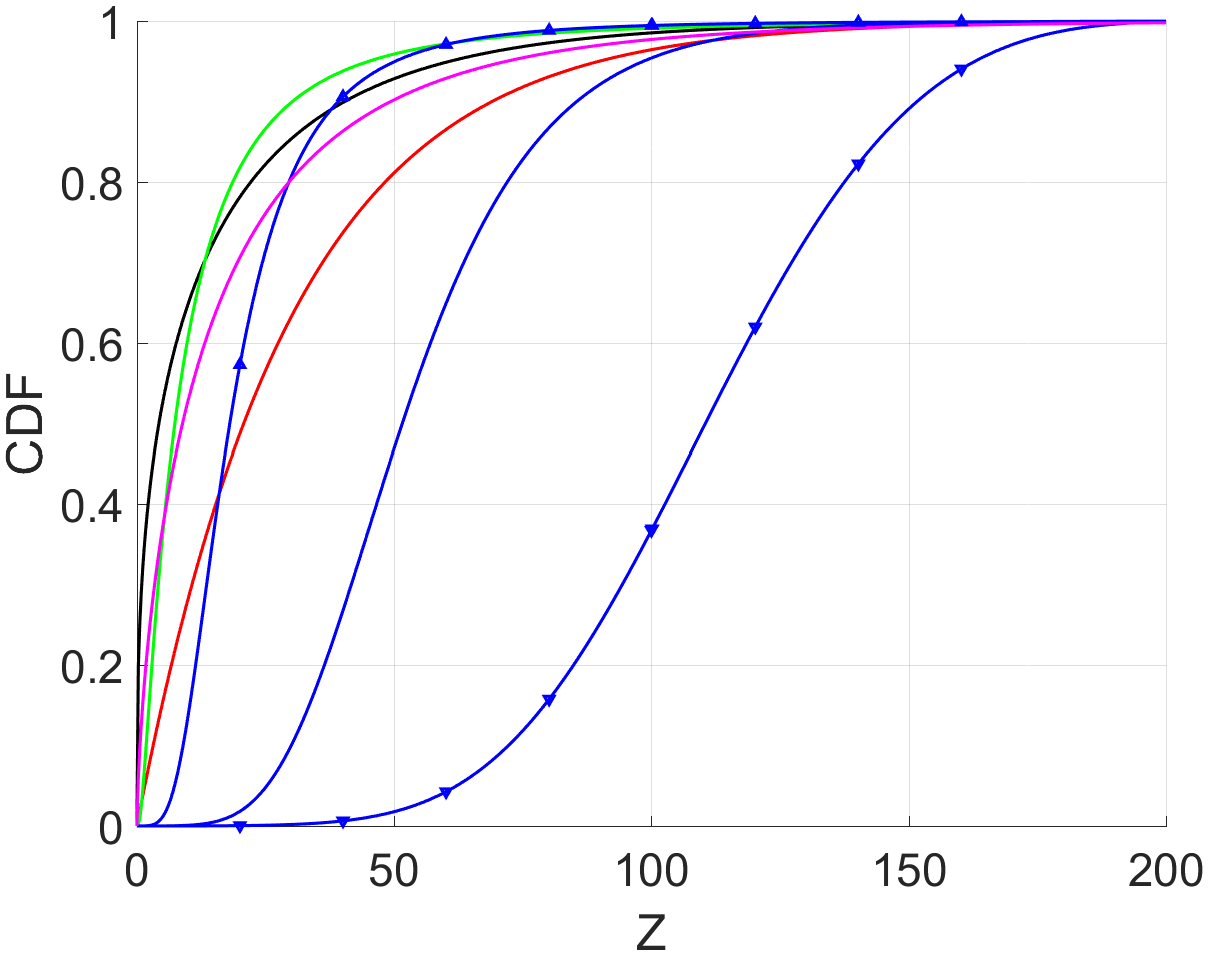}
        \caption{ \label{fig:model-CDF} }
    \end{subfigure}
    \begin{subfigure}[h!]{0.49\textwidth}
        \includegraphics[width=\textwidth]{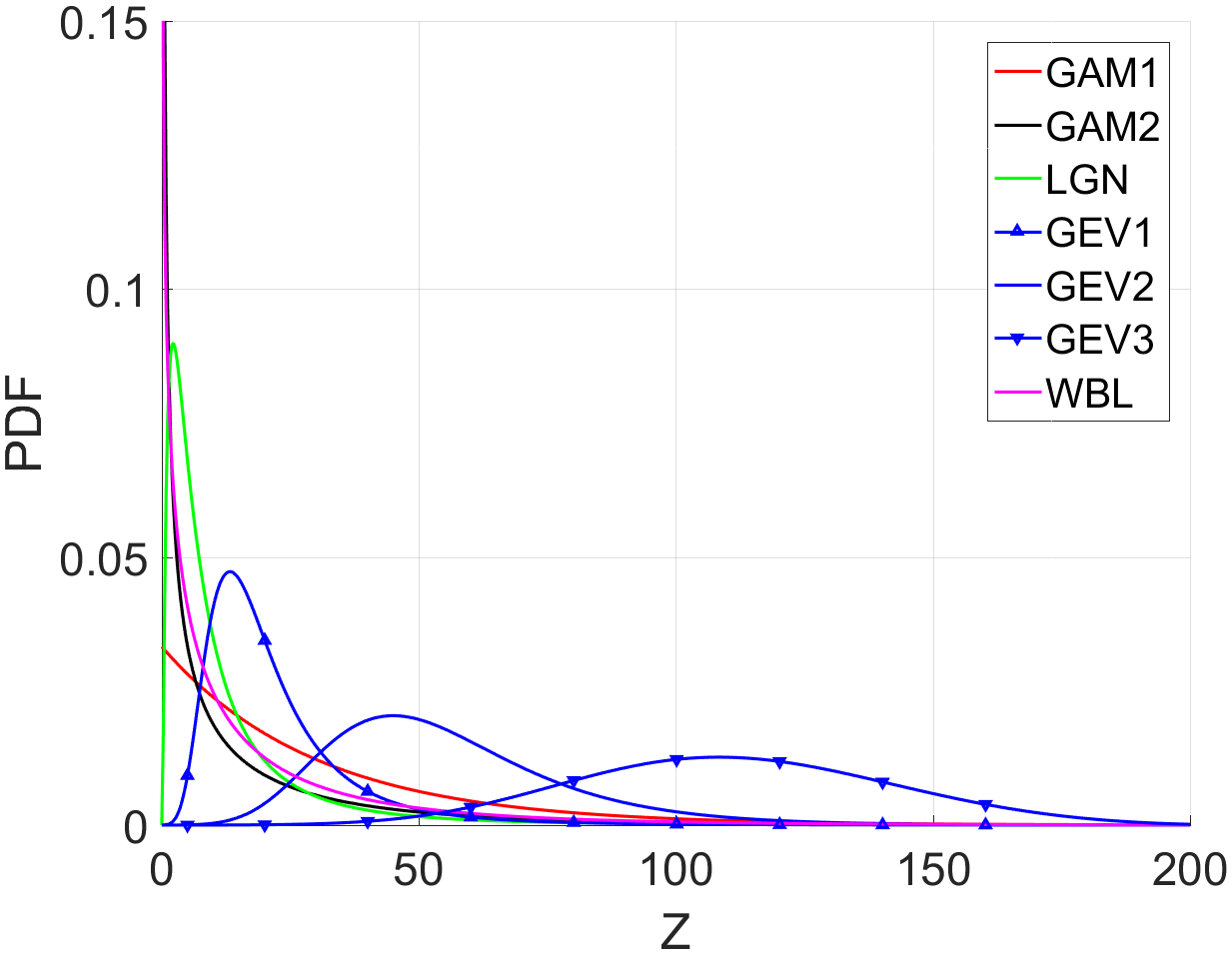}
        \caption{ \label{fig:model-PDF} }
    \end{subfigure}
     \caption{CDF (left) and PDF (right) plots for the six investigated probability models [see Eq.~\eqref{eq:model-cdf}]. GAM1: GAM($\xi$ = 1, $\sigma$ = 30), GAM2: GAM($\xi=0.35$, $\sigma = 40$), LGN: LGN($m=2$, $\sigma=1.1$), GEV1: GEV($\xi=0.25$, $\sigma=8$, $\mu=15$), GEV2: GEV($\xi=0$, $\sigma=18$, $\mu=45$), GEV3: GEV($\xi=-0.25$, $\sigma=30$, $\mu=100$), WBL: WBL($\sigma = 15, \xi = 0.7 $).  The horizontal axis represents monthly precipitation amount (mm).  The parameters $\xi, \sigma, \mu, \lambda, \kappa $ for each model are defined with reference to  Eq.~\eqref{eq:model-pdf}. }
     \label{fig:model-CDFPDF}
 \end{figure}

The studied  probability distributions and their parameters  are: GAM($\xi$ = 1,  $\sigma$ = 30), GAM($\xi=0.35$, $\sigma = 40$), LGN($m=2$, $\sigma=1.1$), GEV($\xi=0.25$, $\sigma=8$, $\mu=15$), GEV($\xi=0$, $\sigma=18$, $\mu=45$), GEV($\xi=-0.25$, $\sigma=30$, $\mu=100$),  and WBL($\sigma = 15, \xi = 0.7 $). The respective theoretical CDFs and PDFs are shown in Fig.~\ref{fig:model-CDFPDF}.

 \begin{figure}[!ht]
\centering
\begin{subfigure}[h!]{0.42\textwidth}
\includegraphics[width=\textwidth]{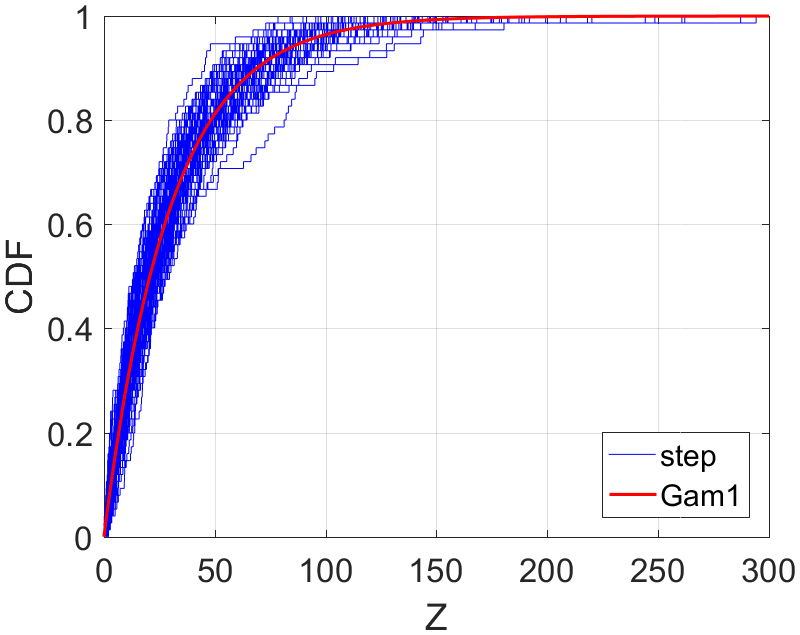}
\end{subfigure}
\begin{subfigure}[h!]{0.42\textwidth}
\includegraphics[width=\textwidth]{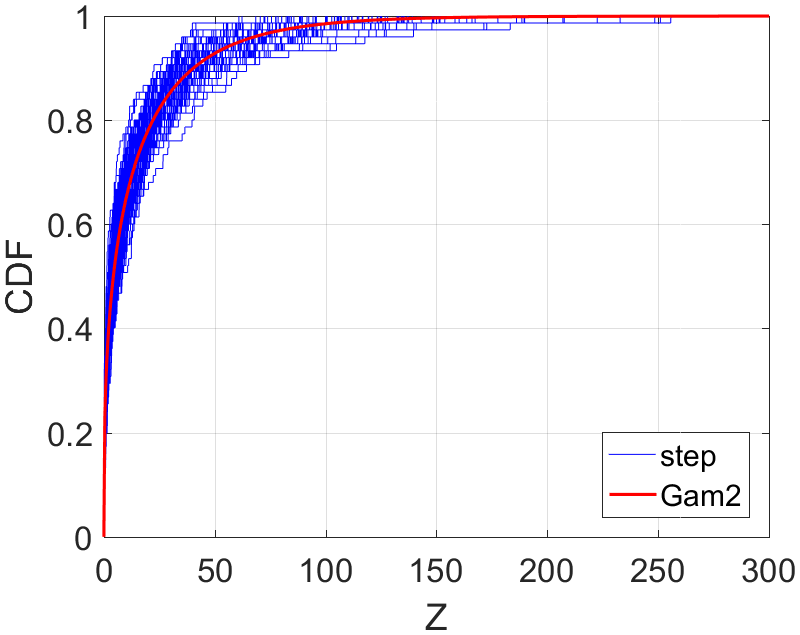}
\end{subfigure}
\begin{subfigure}[h!]{0.42\textwidth}
\includegraphics[width=\textwidth]{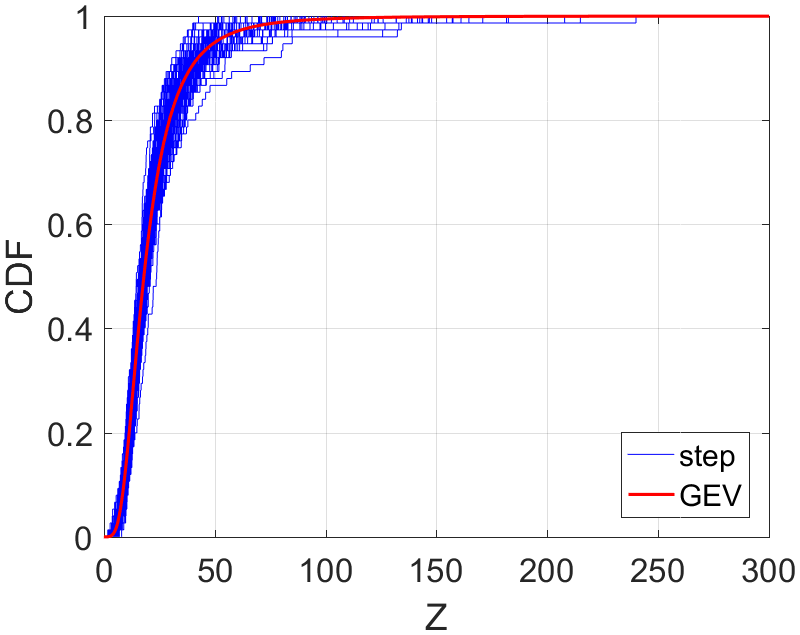}
\end{subfigure}
\begin{subfigure}[h!]{0.42\textwidth}
\includegraphics[width=\textwidth]{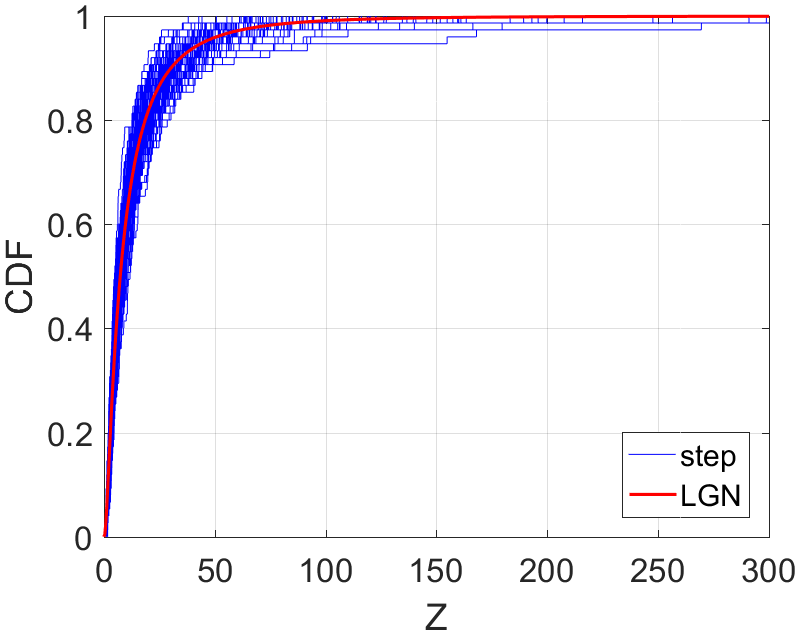}
\end{subfigure}

\caption{Cloud of 100 empirical (staircase) CDFs for four probability distribution models.  GAM1: GAM($\xi$ = 1, $\sigma$ = 30), GAM2: GAM($\xi=0.35$, $\sigma = 40$), GEV1: GEV($\xi=0.25$, $\sigma=8$, $\mu=15$), LGN: LGN($m=2$, $\sigma=1.1$). The CDF estimates are based on samples of length $N=75$. The theoretical CDF curves are shown by the  continuous line (red online). }
\label{fig:Step_func}
\end{figure}

\subsubsection{Comparisons of CDF estimates}
\label{ssec:CDF_compar}

The simulated precipitation amounts comprise the vectors
$Z^{(m)}_{N;j}$ of length   $N \in \{50, 100, 200, 500, 1000\}$,  $m=1, \ldots, M$ $(M=100)$ denotes the sample index, and $j=1, \ldots, 7$ marks the probability distribution model.

The CDF is estimated using the KCDE method with the BGK bandwidth (see Section~\ref{sec:methods}). KCDE is performed with several compactly supported kernels (triweight, uniform, spherical, bitriangular, Epanechnikov) and the infinite-support Gaussian kernel.
For comparison purposes, in addition to the KCDE, $\Fk(z)$, we  calculate the staircase CDF estimate, $\hat{F}_s(z)$, and the theoretical CDF, $F(z)$, at the test  points (times) $t_{k; N,j,m}$, where $k=1, \ldots, N_{t}$ $(N_{t}=500)$. The times $t_{k; N,j,m}$ are uniformly distributed between the  minimum and maximum values for each sample vector  $Z^{(m)}_{N;j}$.  In the following, for ease of notation  we  drop the indices $\{N,j,m\}$ whenever possible.

The mean square error ($\mathrm{MSE}_{k}$) between $\Fk(\cdot)$ and the theoretical CDF $F(\cdot)$ is calculated for all the studied kernels and all the samples $Z^{(m)}_{N;j}$ based on the validation values at the test points.
Similarly, the mean square error ($\mathrm{MSE}_{s}$)  of the staircase function $\hat{F}_s(\cdot)$ with respect to  $F(\cdot)$ is also calculated.

\subsubsection{Assessing KCDE performance}
\label{ssec:KDCE_Comparison_Staircase}
\begin{figure}[!h]
  \centering
    \begin{subfigure}[h!]{0.49\textwidth}
        \includegraphics[width=\textwidth]{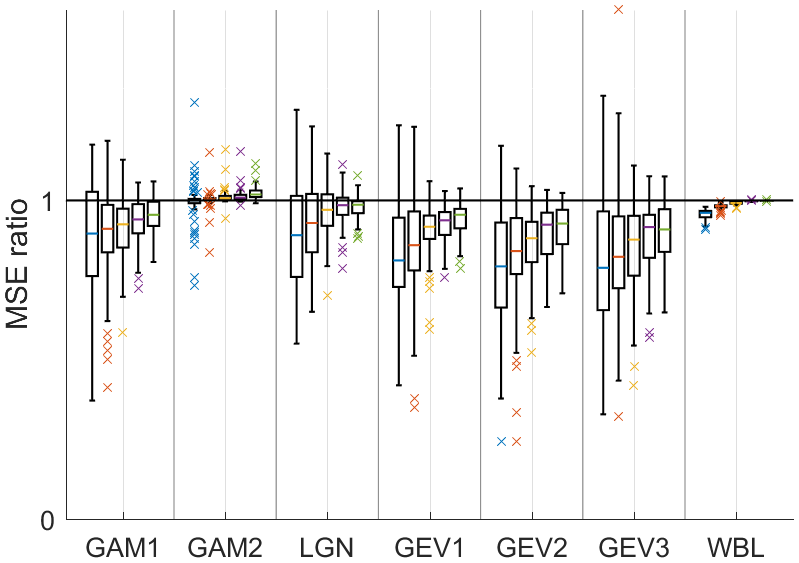}
        \vspace{-20pt}
        \caption{ Bitriangular \label{ fig:RE_Epan } }
    \end{subfigure}
    \begin{subfigure}[h!]{0.49\textwidth}
        \includegraphics[width=\textwidth]{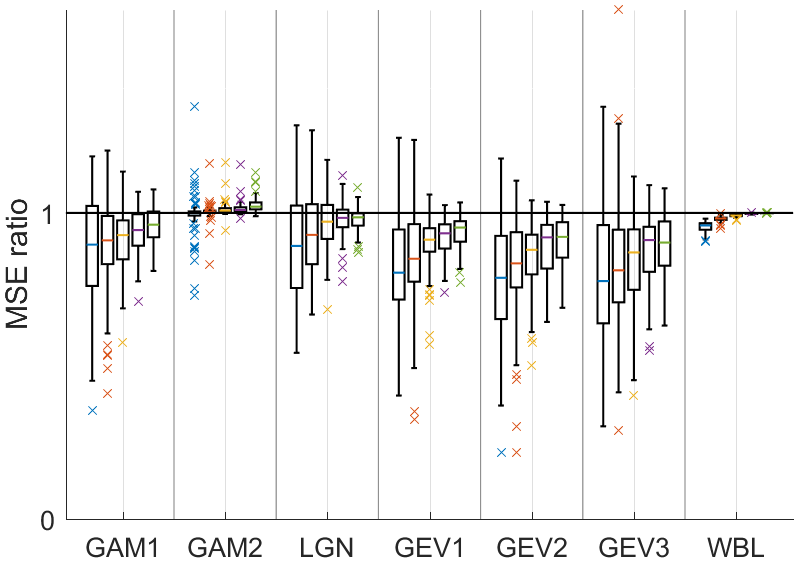}
        \vspace{-20pt}
        \caption{ Triweight \label{ fig:RE_Triw } }
    \end{subfigure}
    \begin{subfigure}[h!]{0.49\textwidth}
        \includegraphics[width=\textwidth]{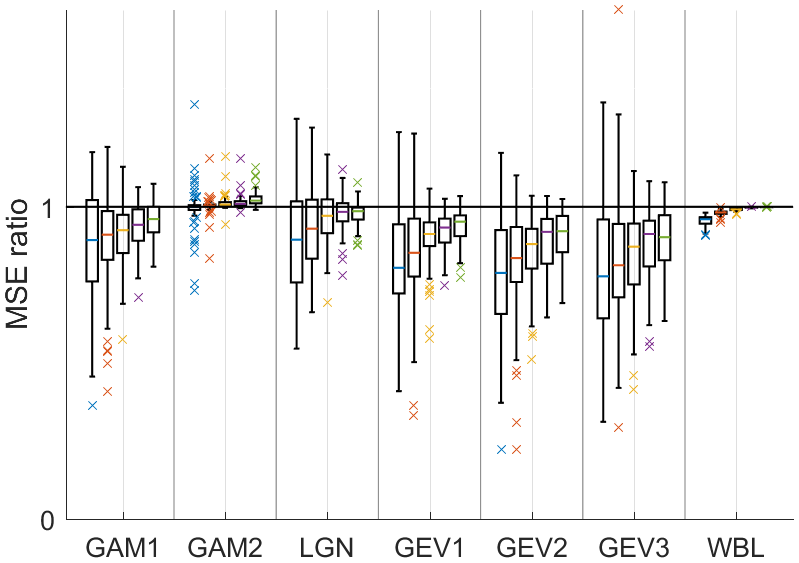}
        \vspace{-20pt}
        \label{ fig:RE_Sphe}
        \caption{ Spherical   }
    \end{subfigure}
    \begin{subfigure}[h!]{0.49\textwidth}
        \includegraphics[width=\textwidth]{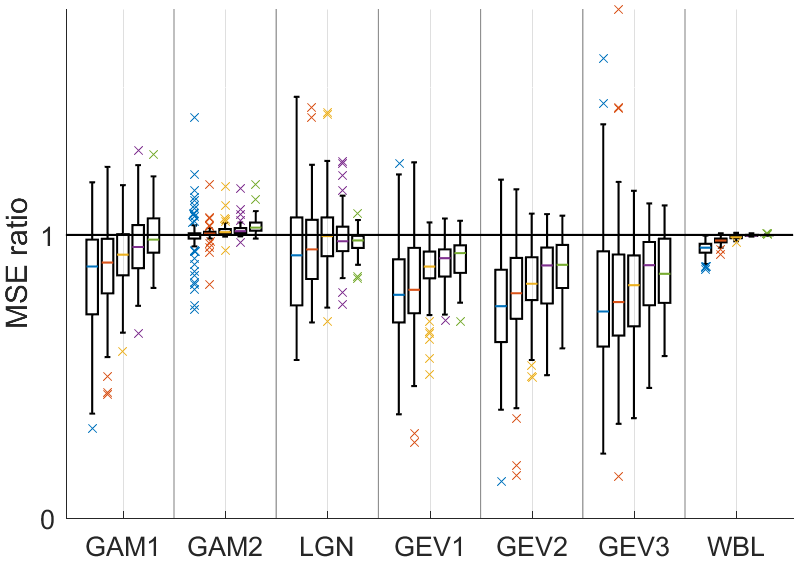}
        \vspace{-20pt}
        \label{ fig:RE_Quad }
        \caption{ Epanechnikov }
    \end{subfigure}
    \begin{subfigure}[h!]{0.49\textwidth}
        \includegraphics[width=\textwidth]{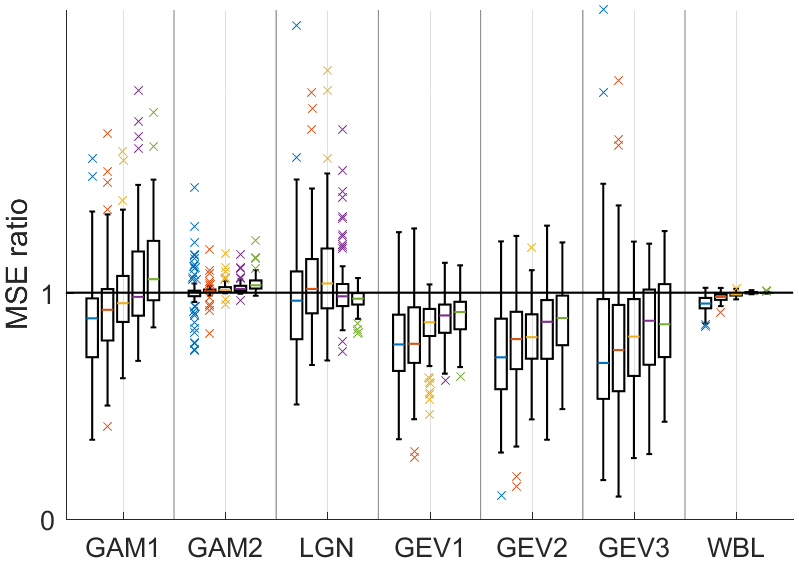}
        \vspace{-20pt}
        \label{ fig:RE_Unif }
        \caption{ Uniform    }
    \end{subfigure}
    \begin{subfigure}[h!]{0.49\textwidth}
        \includegraphics[width=\textwidth]{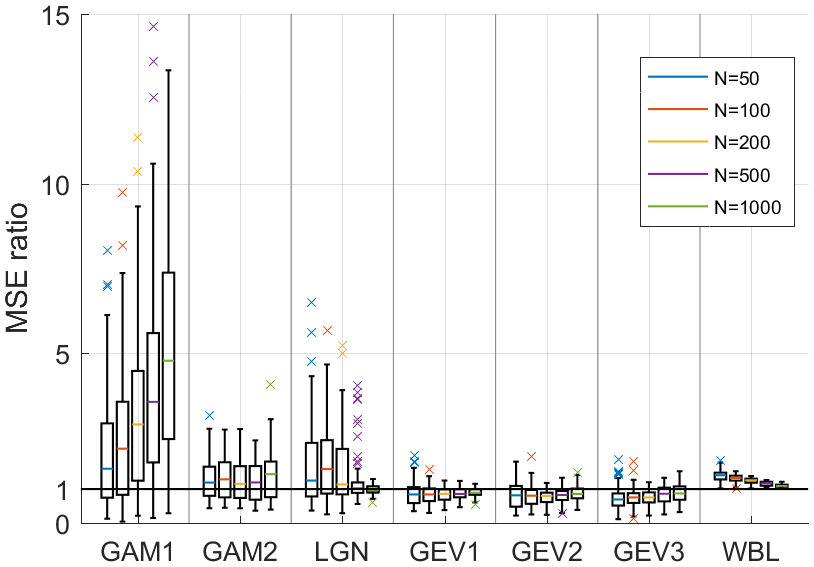}
        \vspace{-20pt}
        \label{ fig:RE_Gaus }
        \caption{ Gaussian }
    \end{subfigure}
    \caption{Boxplots of   RE\textsubscript{k|s}  (MSE of KCDE $\Fk$  divided by the MSE of the staircase  estimate $\hat{F}_s$) for all tested kernels. $N$ is the sample size. The abbreviations of the probability models (GAM1, GAM2, LGN, GEV1, GEV2, GEV3, WBL) are defined in the main text. }
     \label{fig:RE_kernels}
 \end{figure}

The relative performance of  KCDE and the staircase estimate is measured  by means of the error ratio RE\textsubscript{k|s}=$\frac{\mathrm{MSE}_{k} }{\mathrm{MSE}_{s} }$. If RE\textsubscript{k|s}$<1$, the kernel-based CDF $\Fk(\cdot)$ is a better approximation of the theoretical CDF $F(z)$ than the staircase function.
The comparison is illustrated in the box plots of Fig.~\ref{fig:RE_kernels}.

As evidenced in these plots, in most cases the  KCDEs,  $\Fk(z)$, perform better than the staircase estimates $\hat{F}_s(z)$. Hence, $\Fk(z)$, especially for compact kernels, better approximates the CDF of non-Gaussian, skewed distributions than the staircase CDF $\hat{F}_s(z)$. The exception is GAM($\xi=0.35$, $\sigma = 40$), the PDF $f(z)$ of which exhibits a  singularity at zero (see Fig.~\ref{fig:model-CDFPDF}). This violates the BGK smoothness conditions which require the existence of  $f'(z)$ and $f''(z)$.
Based on  Fig.~\ref{fig:RE_kernels},  no single kernel is universally the top performer. The compactly supported kernels tend to outperform the Gaussian kernel. However, in certain cases the Gaussian kernel performs better than all the compact kernels, particularly for  the GEV distribution.

The bitriangular and uniform kernels performed best among the compact kernels, with Bitriangular returning the smallest RE\textsubscript{k|s}.
The spherical kernel, while performing well compared to the step function (see  Fig.~\ref{fig:RE_kernels}), outperforms the other kernels in less than 1\% of the  experiments. However, the deviation of the spherical from  other compact kernels in terms of MSE is within 2\% (based on the average MSE over the 35 different sampling scenarios). In most cases in which the spherical kernel performs well, a different kernel performs slightly better (often the Bitriangular kernel).

The same analysis was performed using the plug-in bandwidth $r^{(m)}_{N;j}$ based on the normal reference rule:

\begin{equation}
\label{h_RoT}
r^{(m)}_{N;j} = 1.06 \sigma^{(m)}_j N^{-1/5}, \; m=1, \ldots M, \; j=1, \ldots, 6,
\end{equation}
where $\sigma^{(m)}_j$ is the standard deviation of the sample $Z^{(m)}_{N;j}$~\cite{scott2015}. The results (Fig.~\ref{fig:RoT_Kern} in~\ref{appC}) show that $\Fk$ using the normal reference bandwidth performs similarly to the staircase estimate $\hat{F}_s$ for the compact kernels. The BGK bandwidth $h^{(m)}_{N;j}$ outperforms the normal reference rule  according to the  error ratio RE\textsubscript{k|s}. The results support the conclusions reached in~\cite{Botev10}.

The  above analysis shows that  KCDE equipped with the   BGK bandwidth  estimates the CDF of synthetic data more reliably than KCDE with the normal reference bandwidth and the empirical staircase estimators. The relative performance of different  kernel functions on the estimate of the CDF $\Fk(z)$ should be further investigated.

\subsubsection{KCDE-ITS simulations}

One simulated state was generated for each probability model in  Section~\ref{ssec:Distributions},  based on  a single sample of length $N=250$.
The simulated time series involve  $N'=250$  points. The CDF $\hat{F_k}(\cdot)$ was estimated using Eq.~\eqref{eq:CDF_Kernel} with the spherical kernel (which has similar performance as the Bitriangular kernel; see Section~\ref{ssec:KDCE_Comparison_Staircase})

\begin{figure}[!h]
  \centering
\includegraphics[width=0.95\textwidth]{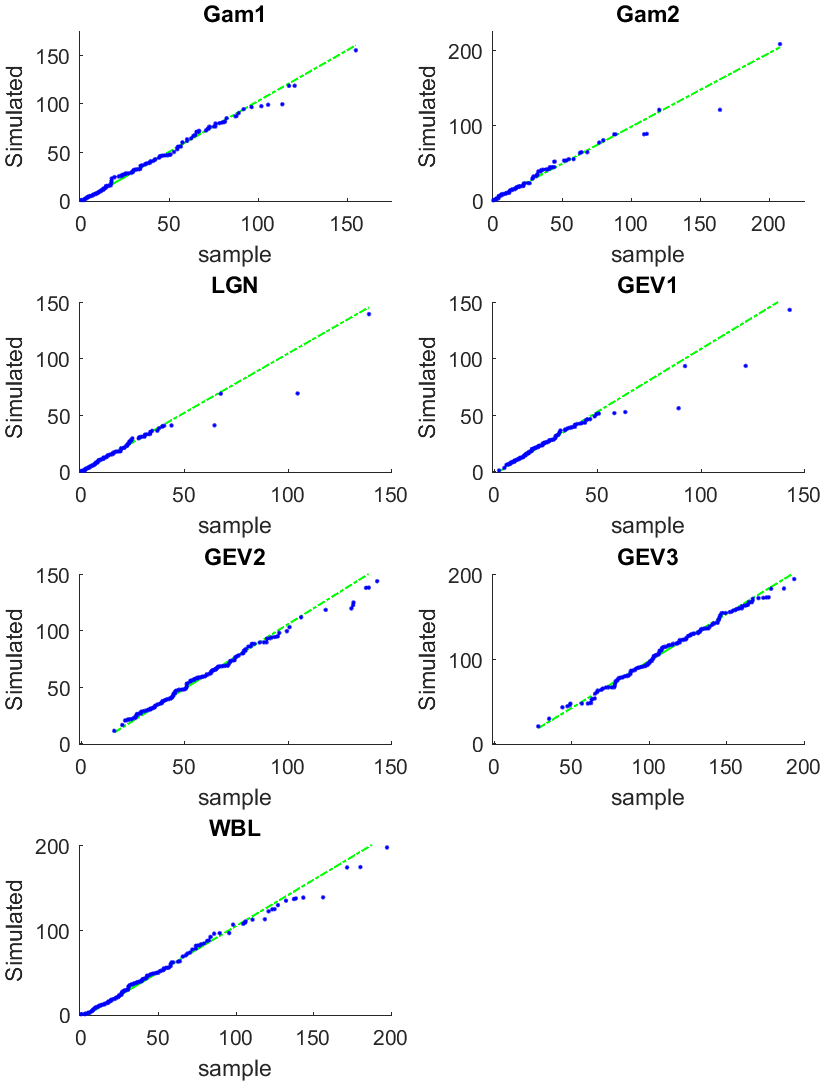}
\caption{Quantile-quantile (q-q) plots between synthetic sample data and ITS simulated sequences of the same length.  The kernel-based  CDF estimate $\hat{F}_k$ was constructed using the spherical kernel. Each synthetic sample includes 250 values drawn from the parametric probability distributions presented in Section~\ref{ssec:Distributions}. }
\label{fig:TS_sim}
 \end{figure}

Figure~\ref{fig:TS_sim} shows the quantile-quantile (q-q) plots between the simulated series and the samples.
The q-q pairs  cluster around the diagonal line, indicating good agreement between the probability distributions of the sample and simulated state, except for a few points in the right tail of the distributions. This behavior signifies underestimation of the maxima by the ITS simulated time series. Given the  skewness of the probability models, the bias stems from the fact that obtaining a simulated sample which contains a value close to the sample maximum  requires generating a uniform random number close to one. This is more likely to happen for  longer simulated series. Note that values higher than the sample maximum can also be obtained by ITS simulation and are more likely for high values of $N'$ compared to $N$ (this case is not shown herein). For compactly supported kernels such values do not exceed the limit $z_{\max}+h$.

\subsection{KCDE application to ERA5 reanalysis data}
\label{ssec:era5}
\begin{figure}[h!]
\centering
\includegraphics[width=0.95\textwidth]{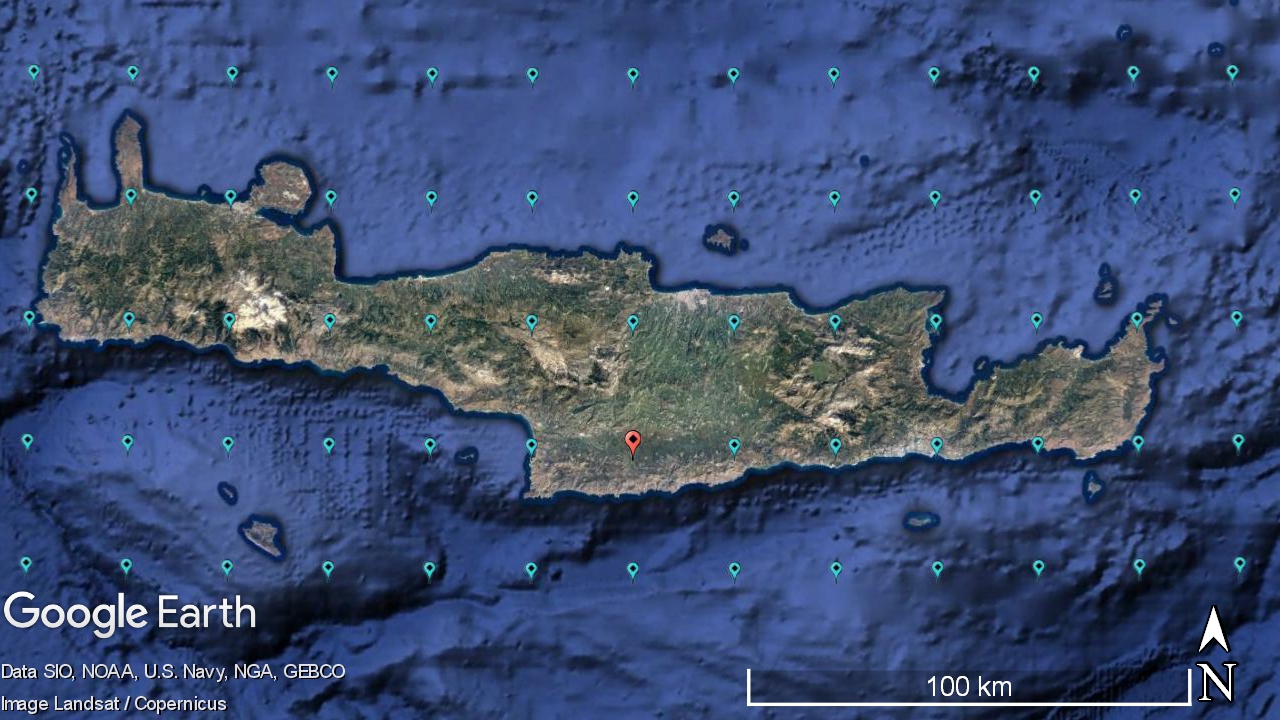} 
\caption{Geomorphological map of Crete showing the  65 ERA5 grid locations (blue markers) on and around the  island of Crete (Greece) used in this study \cite{GoogleEarth}. The node on the fourth row and seventh column (red marker online) is the test location near the Messara valley.}
 \label{fig:ERA5_Grid}
\end{figure}

This section employs ERA5 reanalysis precipitation data (mm) downloaded from the Copernicus Climate Change Service~\cite{copernicus18}. The data set includes $23{,}360{,}610$ values of hourly precipitation amounts for a period of 41 years (from 01-Jan-1979 06:00:00 to 31-Dec-2019 23:00:00) at the nodes of a $5 \times 13$ spatial grid (see Fig.~\ref{fig:ERA5_Grid}) on and around  the island of Crete (Greece).  The average spatial resolution is $\approx0.28$ degrees (grid cell size $\approx 31$km), and  $359{,}394$ hourly precipitation values are available at every node.

Our analysis focuses on  the wet period, i.e., from October to March~\cite{Nastos09}. Daily, weekly, monthly and annual precipitation sets were generated by aggregating the hourly values at each location over the respective time window.  We kept only non-zero precipitation values for our analysis. Note that we also tested replacing zeros  with small values as  done in some  studies~\cite{Wilks90,Wang2010,Tang2021}. Such replacement does not significantly affect the shape of the distributions at any time scale.  The optimal model for the daily data sets (based on selection criteria of Table~\ref{table:best_models_all_nodes_wet_nonzeros})  in certain cases changes from Weibull (if zeros are excluded) to gamma (if zeros are replaced with small values), but in such cases the  two models look quite similar.

\subsubsection{CDF estimation for the ERA5 data}
The KCDE analysis of precipitation data was performed for all  65 grid nodes.  For brevity we mostly discuss results from a single representative node (see Fig.~\ref{fig:ERA5_Grid}) which is located near the major agricultural area of Crete, i.e., the valley of Messara~\cite{Varouchakis13}. The coordinates (LAT, LON) of the test node  in the World Geodetic System (WGS~84) are
35$^\circ$0'0.00"N and  25$^\circ$0'0.00"E. The probability distributions at 5 additional nodes, geographically dispersed over the island, are considered in the simulation study in Section~\ref{ssec:simulations}.

Table~\ref{table:stats_data_N34_dontreplace_wet_nonzeros} presents  summary statistics of wet-period precipitation amounts for the four time scales along with the respective   best-fit parametric models  based on BIC. The optimal parametric models are all non-Gaussian distributions.
More details are shown in
Table~\ref{table:Goddnessoffit_MonthlyReanalysis_N34_wet_nonzeros}  which reports the optimal parametric model  based on  Akaike's Information Criterion (AIC), the Bayesian Information Criterion (BIC), and the negative log likelihood (NLL). The three criteria agree on the optimal distribution models for all time scales considered.
For daily amounts the optimal fit is obtained by the Weibull, followed by the gamma, lognormal, GEV and  the normal distribution.
For weekly precipitation, the  ranking of  the probability distributions (from best to worst) is: gamma, Weibull, lognormal, GEV and normal.
For the monthly data, the  following ranking is obtained: Weibull, gamma, GEV, normal, and lognormal.
Finally, for the annual data the  best model  is the Weibull distribution.

\begin{table}[]
\centering
\caption{Summary statistics for the daily, weekly, monthly and annual precipitation amounts for the wet period (October to March) (from ERA5 reanalysis data)  for the test location on the island of Crete. The coordinates (LAT, LON) of the node are
35$^\circ$0'0.00"N and  25$^\circ$0'0.00"E  in the World Geodetic System (WGS~84).
$N$: Number of records analyzed, $N_{z}$: number of zero records removed, $\overline{z}$: mean value, $z_{0.50}$: median, $z_{\min}$: minimum, $z_{\max}$: maximum, $\sigma_{z}$: standard deviation, cov: coefficient of variation, $s_{z}$: skewness, $k_{z}$: kurtosis. ``Model'' is the best-fit distribution based on the Bayesian Information Criterion (BIC).} 
\label{table:stats_data_N34_dontreplace_wet_nonzeros}
\resizebox{1\textwidth}{!}{%
\renewcommand\tabcolsep{3pt} 
\renewcommand\arraystretch{1} 
\begin{tabular}{ l| c c c c c c c c c c c c}
\toprule
Scale & $N$  & $N_{z}$  &$\overline{z}$ & $z_{0.50}$ & $z_{\min}$ & $z_{\max}$ & $\sigma_{z}$ & cov & $s_{z}$ &\ $k_{z}$ & Model \\
\midrule
\midrule
Daily & 5{,}664 & 1{,}808 & 2.0884  & 0.4377 & 4.7684$\,\times 10^{-4}$ & 56.1037    & 4.4048 & 2.1091  & 4.3233  & 28.6750 & Weibull \\ 

Weekly & 1{,}055 & 13 & 11.2781  & 6.3763 & 4.7684$\,\times 10^{-4}$ & 84.3720    & 12.6180 & 1.1188  & 1.5697  & 5.4716 & gamma\\ 

Monthly & 246 & 0  & 48.0850 & 44.3109 & 0.8783  & 142.1008    & 29.4220 & 0.6119  & 0.6378  & 3.0325  & Weibull \\ 

Annual & 41  & 0 & 288.5103  & 303.1404 & 163.4173 & 445.0781    & 66.7700 & 0.2314  & $-$0.1206  & 2.5123 & Weibull \\ 
\bottomrule
\end{tabular}
}
\end{table}

For completeness, Table~\ref{table:best_models_all_nodes_wet_nonzeros} reports the frequency (total number of ERA5 nodes) with which a parametric model is selected as the optimal probability distribution for each time scale. As in the case of the test location, all three criteria agree on the optimal model  except for  the  annual time scale.
On the daily scale, the Weibull model is selected at 56 nodes, followed by the gamma model at 9 nodes. For the weekly scale the gamma distribution is the optimal model at all 65 nodes. For the monthly scale, the Weibull distribution is the optimal model at most nodes according to all three criteria (the number of nodes differs depending on the criterion).
Finally, for the annual precipitation AIC and BIC select the gamma model at 28 nodes, the normal at 18 nodes, the lognormal at 14 nodes and the Weibull at 5 nodes. For the annual  data,  the Weibull distribution provides the best fit at the test location (and its four nearest nodes in the south), while the gamma distribution is  selected at the majority of the ERA5 sites.

\subsubsection{Estimates of precipitation amount at different time scales}

From the analysis above it  follows that the optimal parametric model depends on both the time scale considered and the geographic location. This variability motivates the use of  non-parametric estimates of the underlying  probability distribution.
This section presents KCDEs of  precipitation amount at the test location over the  daily, weekly, monthly and annual time windows.

\begin{figure} [htb!]
\centering
\begin{subfigure}[h!]{0.49\textwidth}
\includegraphics[width=\textwidth]{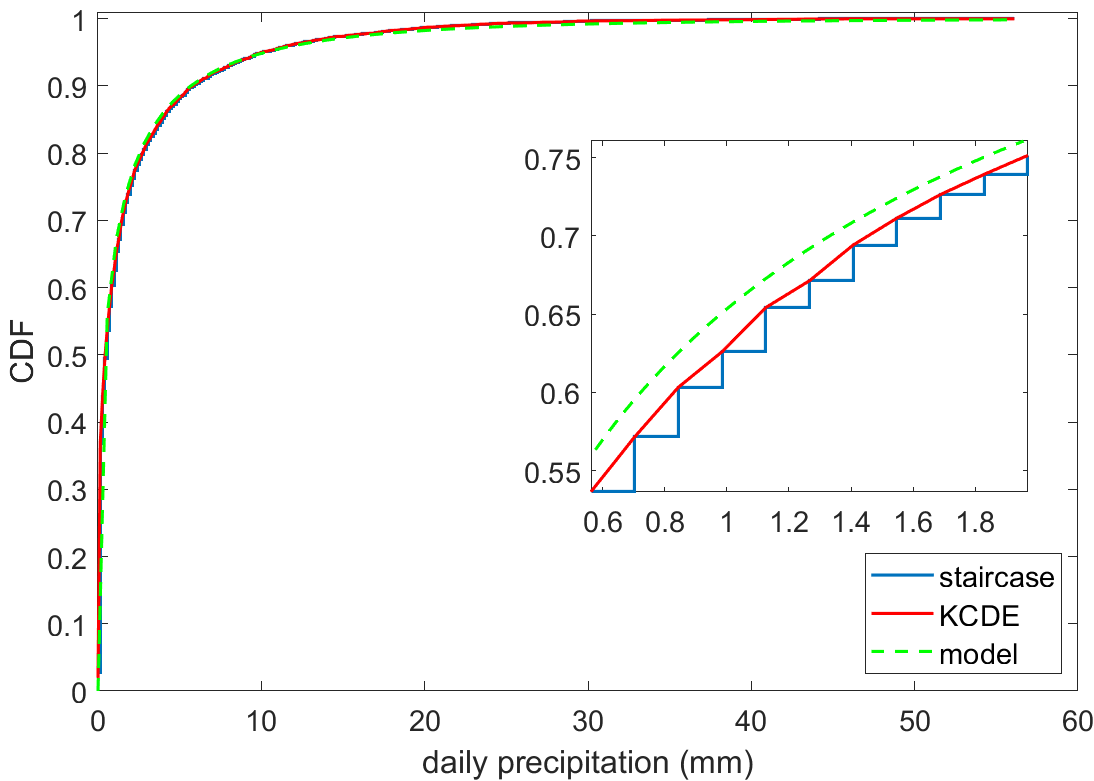}
\label{fig:Dailywet_N34_TriaKern_dontreplace}
\caption{Daily N34 (5{,}664 (from 7{,}472) values)}
\end{subfigure}
\begin{subfigure}[h!]{0.49\textwidth}
\includegraphics[width=\textwidth]{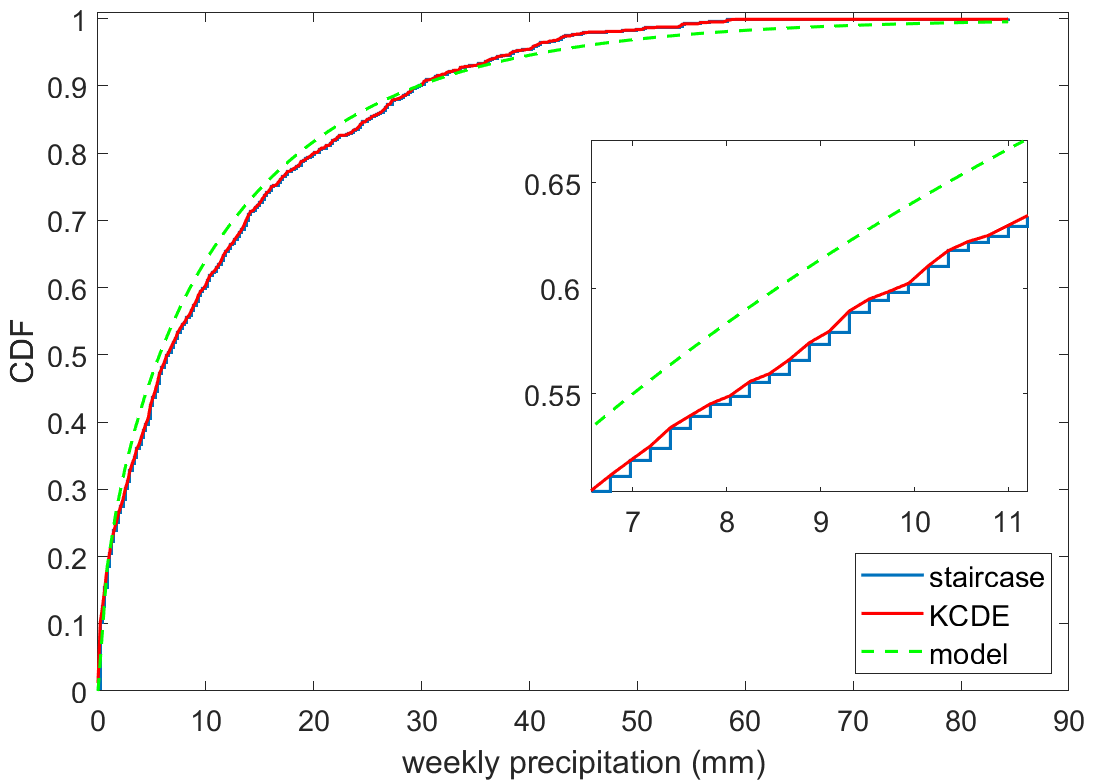}
\label{fig:Weeklywet_N34_TriaKern_dontreplace}
\caption{Weekly N34 (1{,}055 (from 1{,}068) values)}
\end{subfigure}
\begin{subfigure}[h!]{0.49\textwidth}
\includegraphics[width=\textwidth]{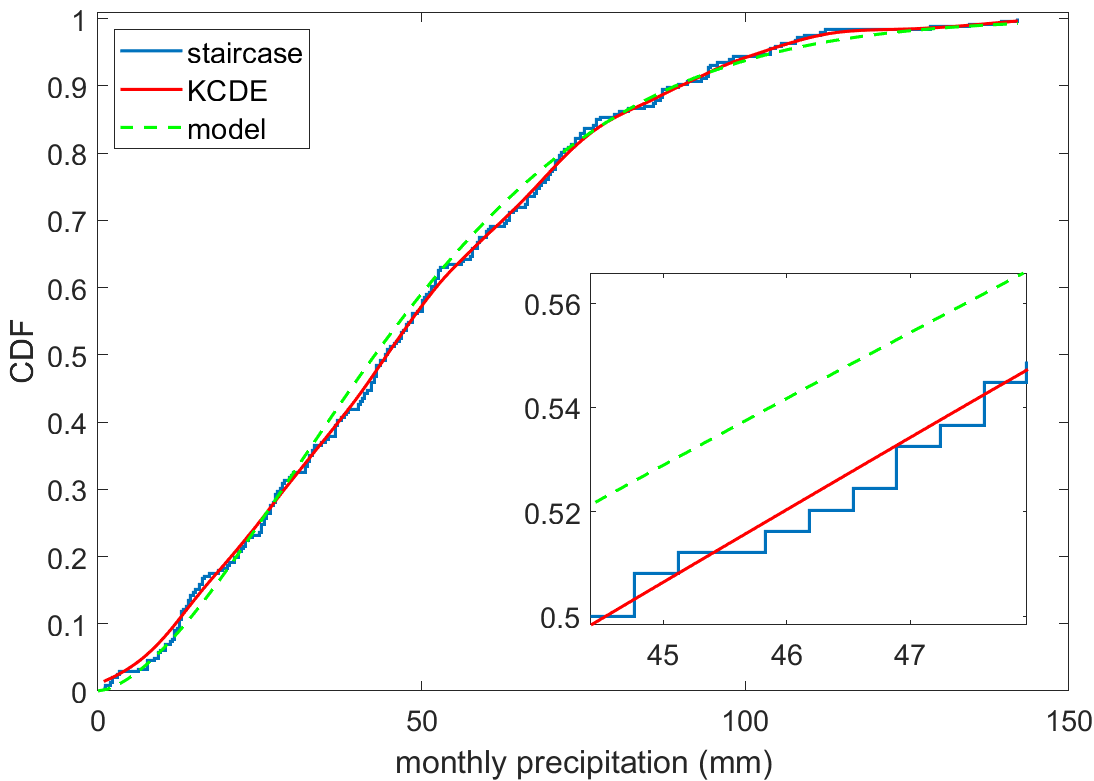}
\label{fig:Monthlywet_N34_TriaKern_dontreplace}
\caption{Monthly N34 (246 values)}
\end{subfigure}
\begin{subfigure}[h!]{0.49\textwidth}
\includegraphics[width=\textwidth]{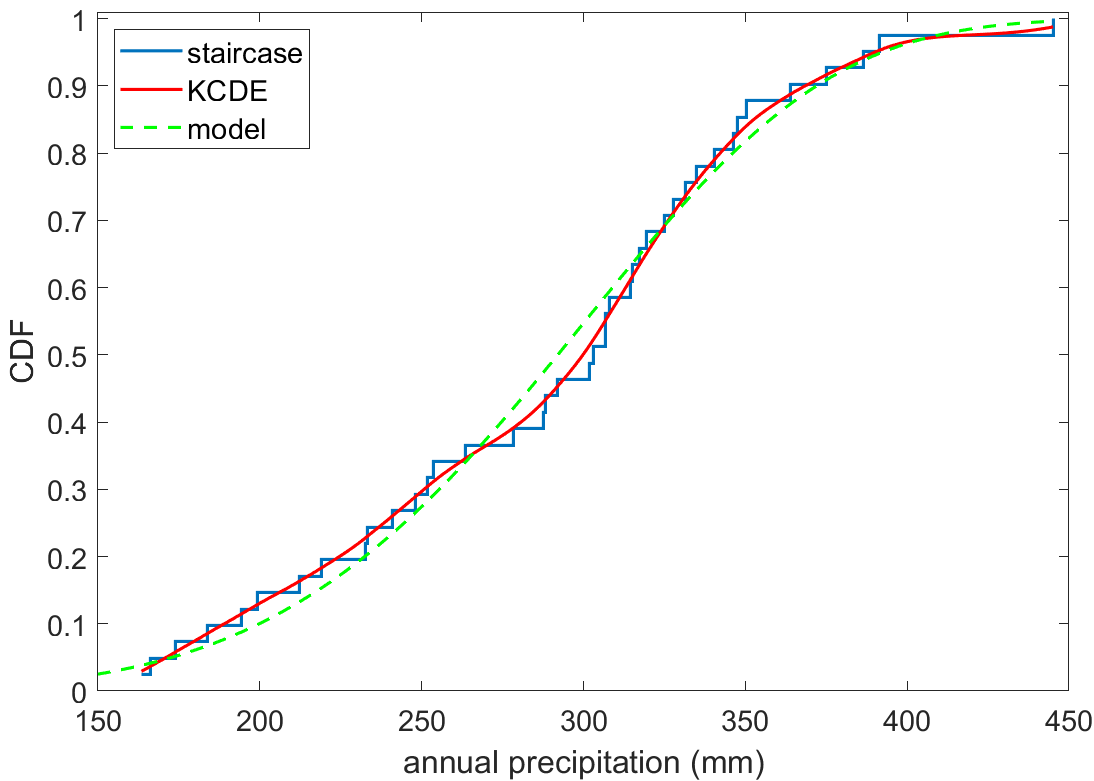}
\label{fig:Annualwet_N34_TriaKern_dontreplace}
\caption{Annual N34 (41 values)}
\end{subfigure}
\caption{Empirical--staircase (step function, blue online) and KCDE with  Bitriangular kernel (smooth line, red online) CDF estimates for the daily (upper left), weekly (upper right), monthly (bottom left) and annual (bottom right) precipitation amount reanalysis data for the wet period (October to March). The dashed line (green online) corresponds to CDFs of the  optimal parametric models based on the Bayesian Information Criterion (BIC). Based on  BIC, the optimal models are Weibull (daily), gamma (weekly), Weibull (monthly), and Weibull (annual) scales respectively.
}
\label{fig:Tria N34 wet period data with dontreplace wet Nonzeros}
\end{figure}

The KCDEs are calculated with the Bitriangular kernel which performed best (among the studied compact kernels) on  synthetic data, as discussed in  Section~\ref{sec:synth_data}.
The sample size for each time window is shown in Table~\ref{table:stats_data_N34_dontreplace_wet_nonzeros}.
Figure~\ref{fig:Tria N34 wet period data with dontreplace wet Nonzeros} compares the empirical (staircase), and KCDE estimates with the best-fit parametric models for each time scale.  As is evident in these plots, the continuous KCDE curves are  closer to the empirical estimates than the  parametric models.  The behavior reflects the fact that parametric models are not flexible enough to capture the natural stochasticity of precipitation.  The differences between the three curves are smallest in the case of the annual data, which are closer to the normal distribution (see Table~\ref{table:best_models_all_nodes_wet_nonzeros}).  The respective BGK bandwidths for the Bitriangular kernel are  $h_d=0.0017$, $h_w=0.027$, $h_m=14.898$, and $h_a=40.824$ (all in mm) for the four different time scales.  These results show that the BGK bandwidth adapts to the shape and scale of the underlying distribution. The inset plots in  Fig.~\ref{fig:Tria N34 wet period data with dontreplace wet Nonzeros} demonstrate that KCDE provides estimates  closer to the empirical CDF  than the optimal parametric models.
Note that the BGK bandwidth for the monthly wet-season data is $h_m=14.898$~mm while the respective bandwidth for the monthly entire-season data set is $h_m=0.1413$~mm (the analysis for the entire season is not reported herein).
This difference is due to the presence of a large number of low values in the full data set and illustrates the ability of the BGK bandwidth to adapt to the data distribution.


\subsubsection{KCDE-ITS simulation study  based on ERA5 reanalysis data}

Wet-period ERA5 daily precipitation data from 1979 to 2019 at six geographically dispersed nodes are used to demonstrate the application of the ITS simulation method.
The selected locations  are listed in Table~\ref{table:TestLocations}.

\begin{table}[!htb]
\centering
\caption{List of the selected ERA5 nodes. The node coordinates (LAT, LON) are  shown in the World Geodetic System (WGS~84). }
\resizebox{1\textwidth}{!}{%
\begin{tabular}{llll}
Index $s$ & Node  & Area Name & (LAT, LON) \\
\hline
$s=1$	& N12	& Chania	& (35$^\circ$30'0.00"~N,  24$^\circ$00'0.00"~E) \\
$s=2$	& N13	& Lefka Ori (White Mountains) & (35$^\circ$15'0.00"~N,  24$^\circ$00'0.00"~E) \\
$s=3$	& N18	& Askifou plateau	& (35$^\circ$15'0.00"~N,  24$^\circ$15'0.00"~E) \\
$s=4$	& N28	& Psiloritis mountain	& (35$^\circ$15'0.00"~N,  24$^\circ$45'0.00"~E) \\
$s=5$	& N34	& Vagionia	& (35$^\circ$00'0.00"~N,  24$^\circ$00'0.00"~E) \\
$s=6$	& N49	& Ierapetra	& (35$^\circ$00'0.00"~N,  25$^\circ$45'0.00"~E) \\
\hline
\end{tabular}
}
\label{table:TestLocations}
\end{table}

The time series $Z_{N;s,t}$, where $t=1,\dots,41$, and $s=1,\dots,6$ denotes the precipitation amount for the wet period  of a specific year (indexed by $t=\mathrm{year}-1978$) at a specific node (denoted by $s$).
For example, the time series $Z_{182;4,1}$   comprises  $N=182$ values corresponding to the wet-period days at location $s=4$ for the year 1979.
Aggregate time series, $Z_{N;s}$, comprise the wet-period data for the entire 41-year window at each location.

\begin{figure} [ht!]
\centering
\begin{subfigure}[h!]{0.45\textwidth}
\includegraphics[width=\textwidth]{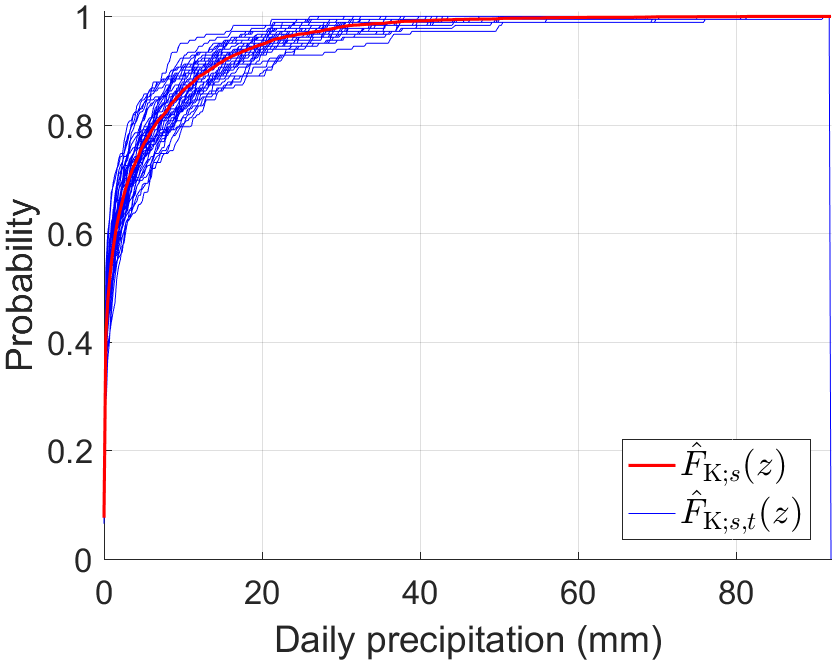}
\caption{Location 1: Chania, N12}
\end{subfigure}
\begin{subfigure}[h!]{0.45\textwidth}
\includegraphics[width=\textwidth]{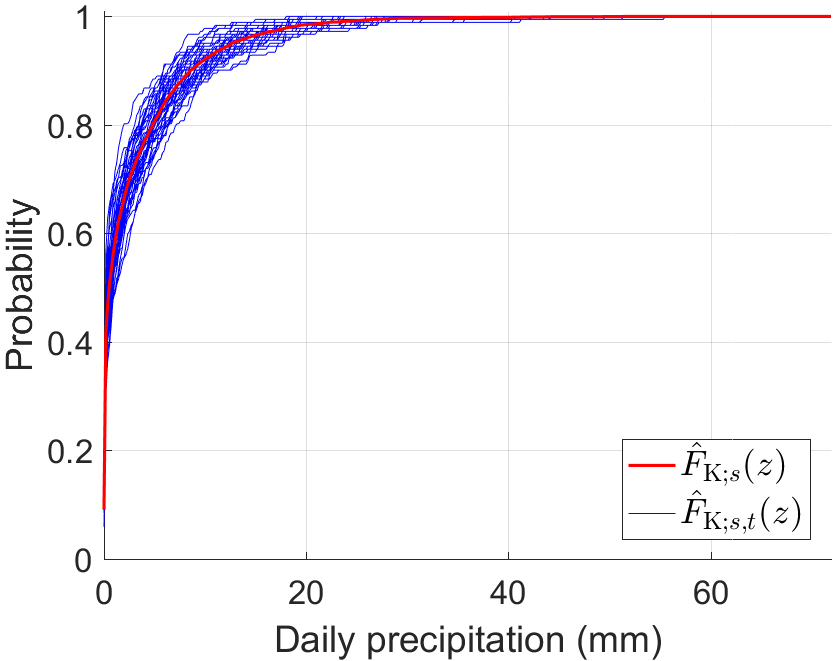}
\caption{Location 2: Lefka Ori, N13}
\end{subfigure}
\begin{subfigure}[h!]{0.45\textwidth}
\includegraphics[width=\textwidth]{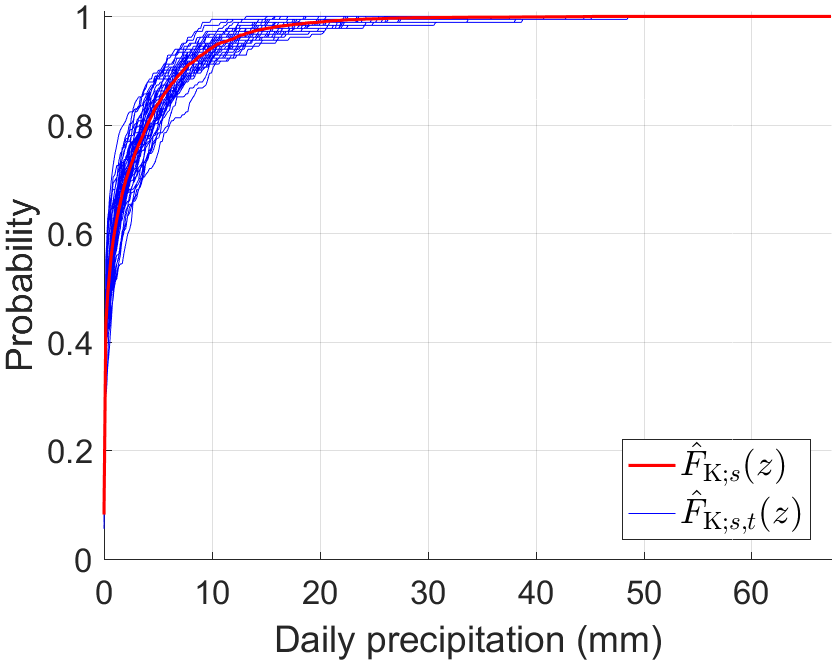}
\caption{Location 3: Askifou, N18}
\end{subfigure}
\begin{subfigure}[h!]{0.45\textwidth}
\includegraphics[width=\textwidth]{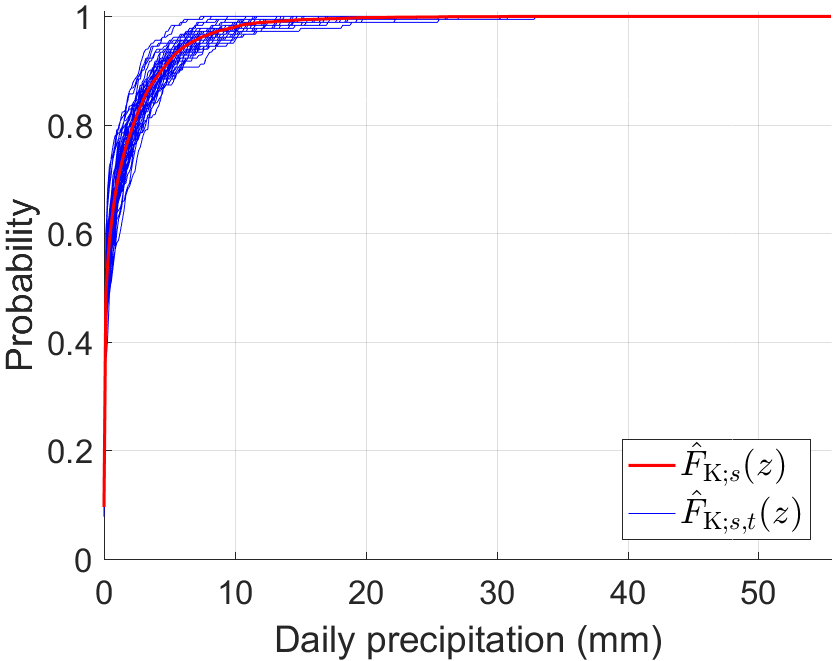}
\caption{Location 4: Psiloritis, N28}
\end{subfigure}
\begin{subfigure}[h!]{0.45\textwidth}
\includegraphics[width=\textwidth]{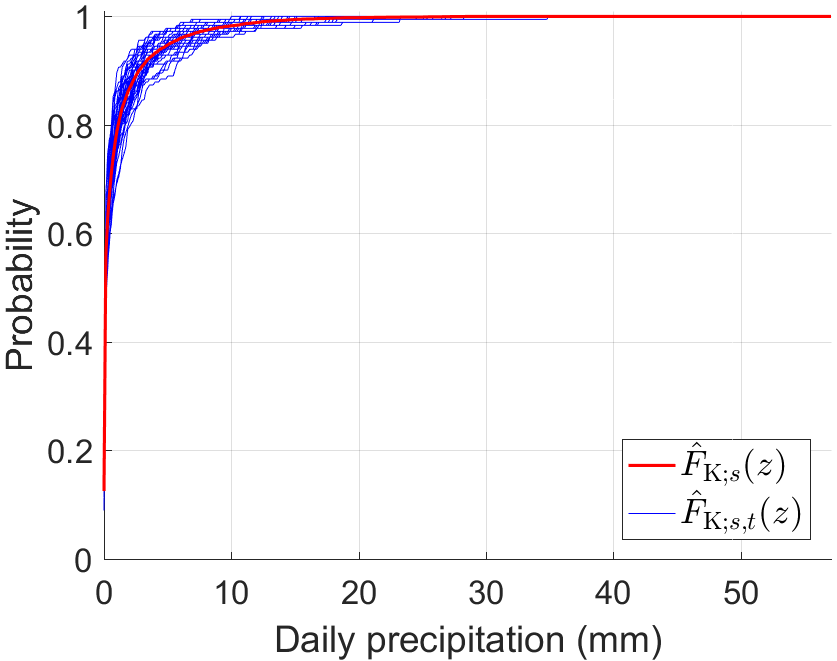}
\caption{Location 5: Vagionia, N34 }
\end{subfigure}
\begin{subfigure}[h!]{0.45\textwidth}
\includegraphics[width=\textwidth]{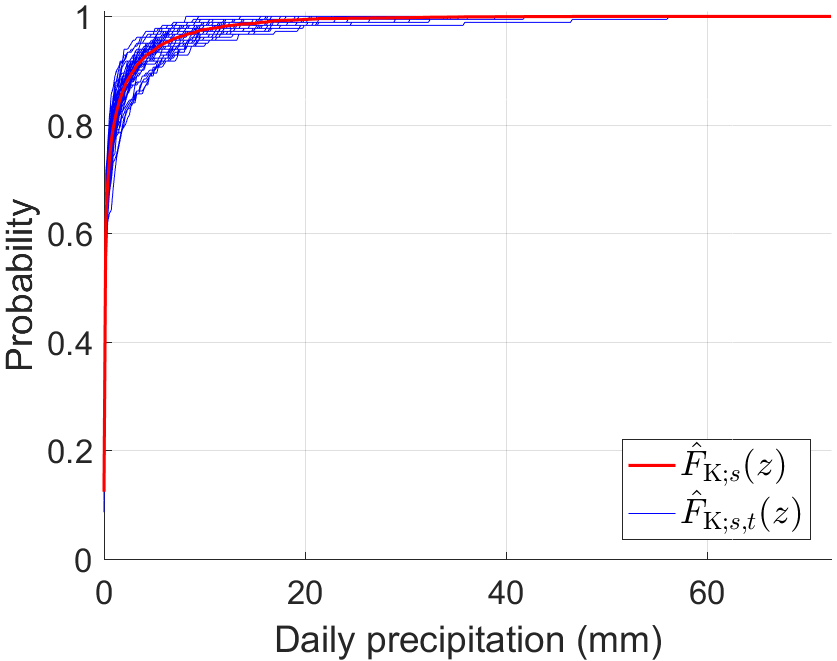}
\caption{Location 6: Ierapetra, N49 }
\end{subfigure}
\caption{CDF $\Fkst(z)$ clouds for  precipitation amount of individual years (blue lines) and $\Fks(z)$ for the aggregate precipitation (red lines) at six ERA5 nodes. The Bitriangular kernel with BGK bandwidth was used for CDF estimation.}
\label{fig:CDFs_WP}
 \end{figure}

The CDF estimates $\Fks(z)$  for $Z_{N;s}$  and $\Fkst(z)$ for $Z_{N;s,t}$, for $s=1,\dots,6$ and $t=1,\dots,41$ were calculated using the Bitriangular kernel with the BGK bandwidth (which are henceforth used). The results are shown in Fig.~\ref{fig:CDFs_WP}.
All CDFs have similar shapes which are quite different from the S-shaped normal CDF. The CDFs exhibit significant probability mass for small values and  extended right tails. The maximum amounts differ slightly between locations.

To test the ITS simulation method we used the following study design.
For each of the years 2015--2019 ($t=37,\ldots, 41$),  the time series $Z_{N;s,t}$ for the years $t=1,\dots,T-1$ were used as the training set. The  precipitation time series $Z_{N;s,T}$ for the  year $t=T$ were considered as the validation set. For each year, an aggregate time series $Z_{N;s}$ was constructed based on the data of the respective training set. Based on the latter, 1000 times series for the  year $T$ were simulated using ITS.
The simulated time series are denoted by  $Z_{N;s,t}^{(m,T)}$, where $t=1,\ldots, T-1, \, m=1, \ldots, 1000$ is the state index, \, $s=1,\dots,6$ and $T=37,\ldots, 41$ is the index of the validation year.

\subsubsection{Analysis of simulation results}
\label{sssec:era5-simulations}

\begin{figure} [ht!]
\centering
\begin{subfigure}[h!]{0.85\textwidth}
\includegraphics[width=\textwidth]{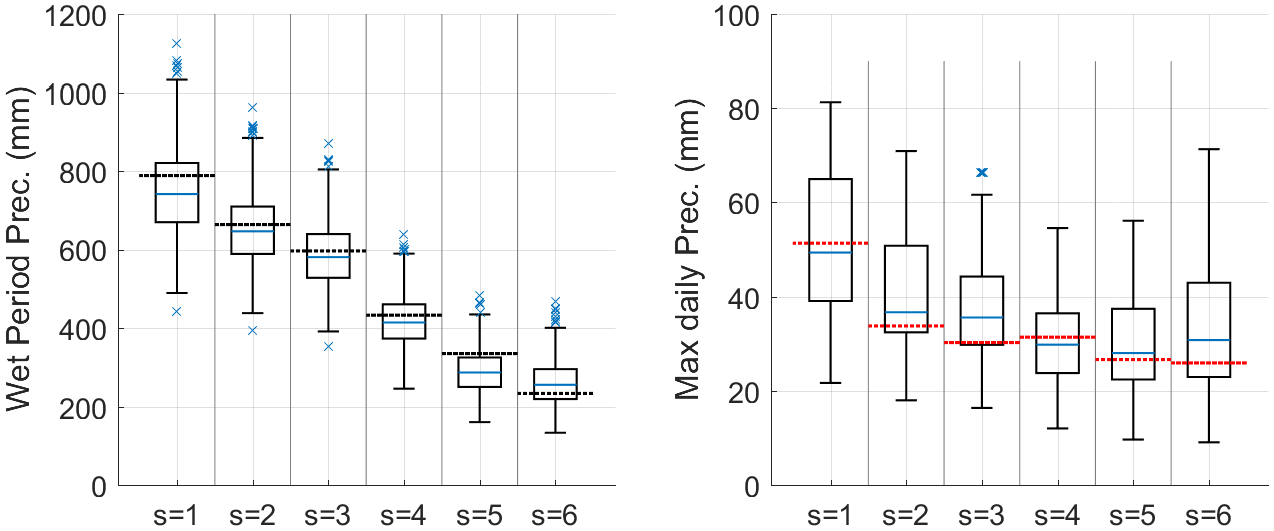}
\caption{Year 2015}
\end{subfigure}
\begin{subfigure}[h!]{0.85\textwidth}
\includegraphics[width=\textwidth]{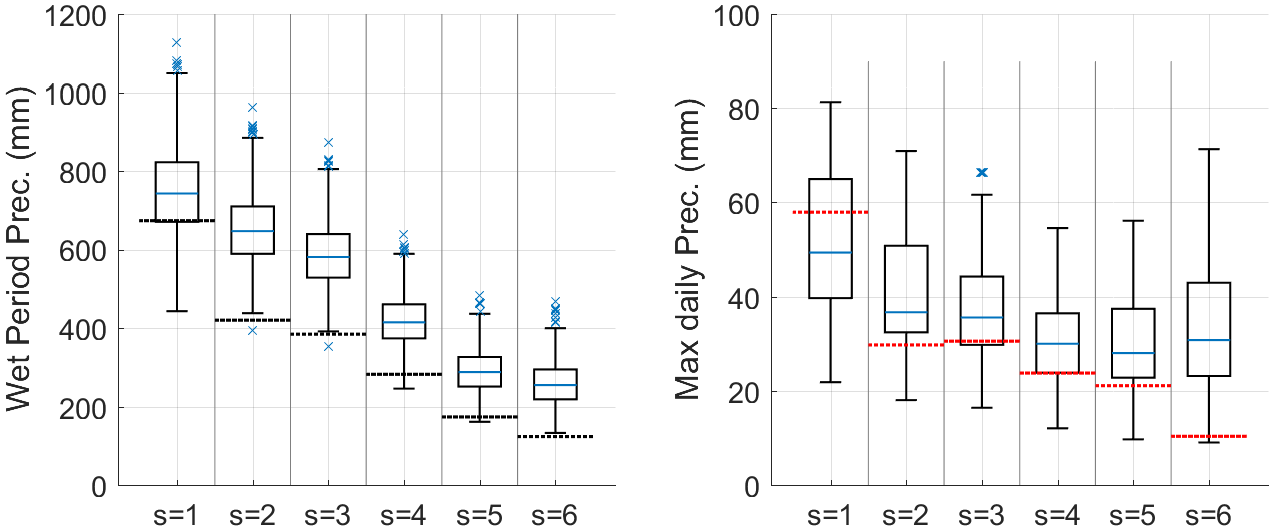}
\caption{Year 2016}
\end{subfigure}
\begin{subfigure}[h!]{0.85\textwidth}
\includegraphics[width=\textwidth]{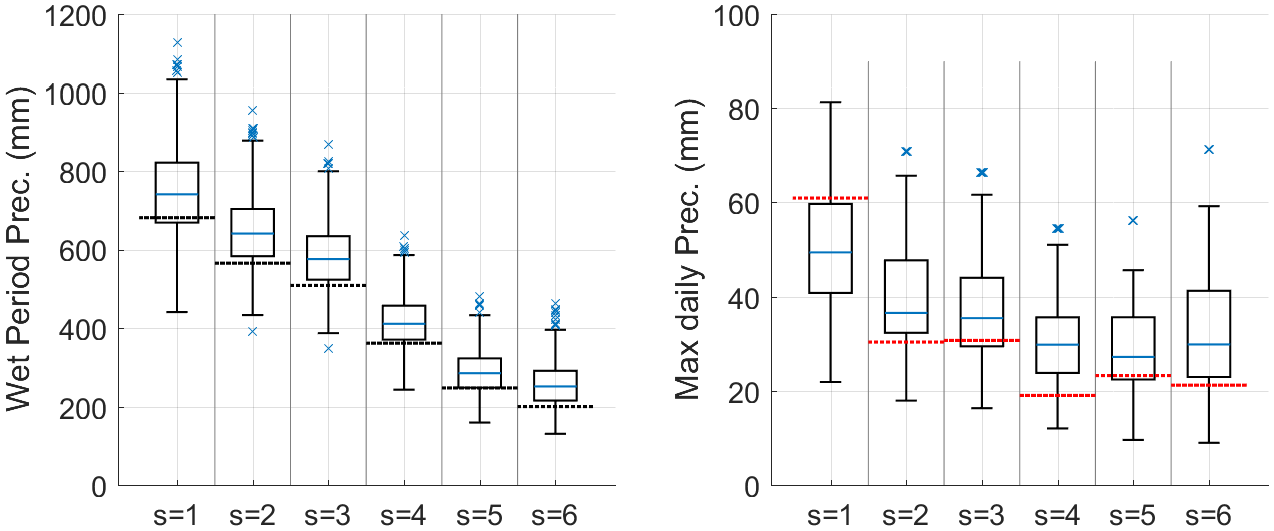}
\caption{Year 2017}
\end{subfigure}
\caption{Boxplots of wet-period total (left column) and maximum daily (right column) precipitation based on 1000 ITS simulations at six ERA5 nodes: $s_{1}$:  Chania, $s_{2}$: Lefka Ori, $s_{3}$: Askifou, $s_{4}:$ Psiloritis, $s_{5}$: Vagionia, $s_{6}$: Ierapetra. Dotted horizontal lines extending outside the boxes represent the  total (black, left column) and  maximum daily (red, right column) ERA5  precipitation values for the validation years 2015-2017. }
\label{fig:Boxplots_WP1}
 \end{figure}

\begin{figure} [ht!]
\centering
\begin{subfigure}[h!]{0.85\textwidth}
\includegraphics[width=\textwidth]{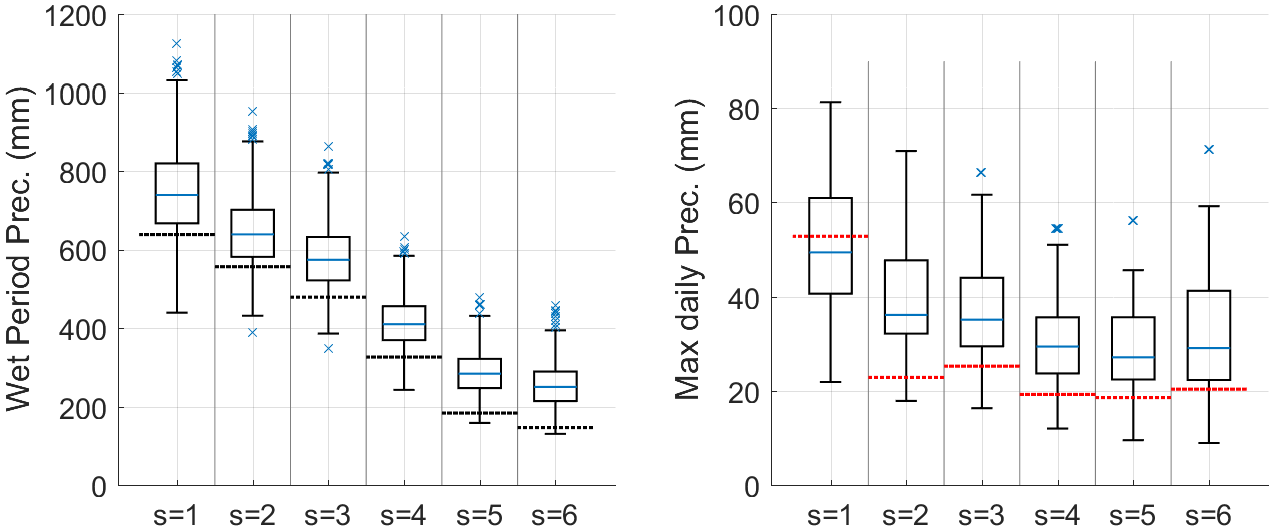}
\caption{Year 2018}
\end{subfigure}
\begin{subfigure}[h!]{0.85\textwidth}
\includegraphics[width=\textwidth]{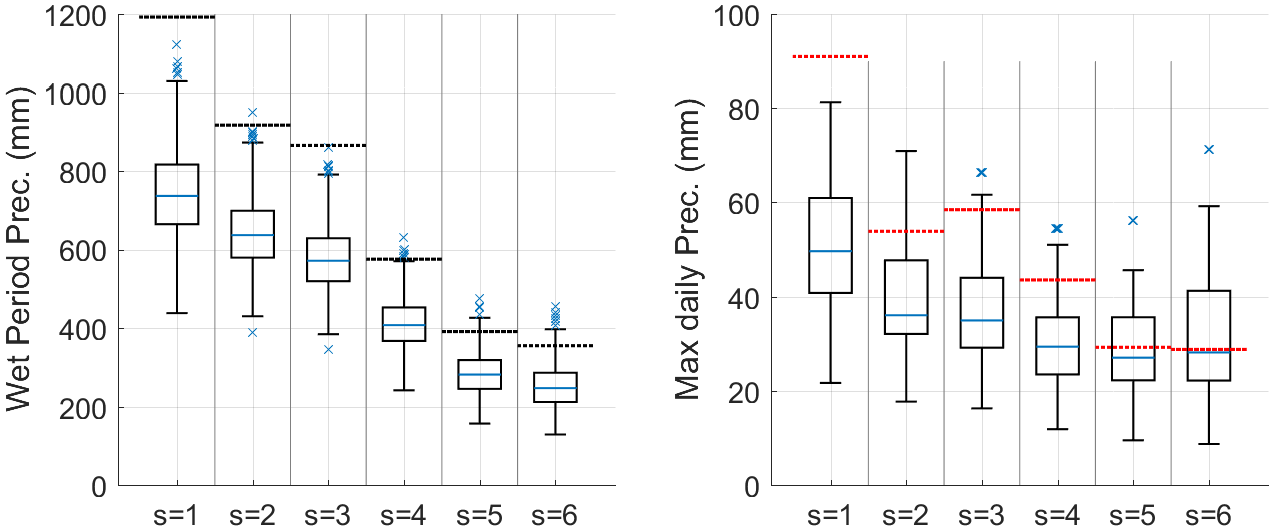}
\caption{Year 2019}
\end{subfigure}
\caption{Boxplots of wet-period total (left column) and maximum daily (right column) precipitation based on 1000 ITS simulations at six ERA5 nodes: $s_{1}$:  Chania, $s_{2}$: Lefka Ori, $s_{3}$: Askifou, $s_{4}:$ Psiloritis, $s_{5}$: Vagionia, $s_{6}$: Ierapetra. Dotted horizontal lines extending outside the boxes represent the  total (black, left column) and  maximum daily (red, right column) ERA5  precipitation values for the validation years 2018-2019.  }
\label{fig:Boxplots_WP2}
 \end{figure}

Figures~\ref{fig:Boxplots_WP1}-\ref{fig:Boxplots_WP2} present boxplots of total precipitation during the  wet period of the validation year (on the left column) as well as the maximum daily precipitation (on the right column). These boxplots are based on the 1000 simulated states for each node $s$ and validation year $T$.
The respective wet-period total and maximum daily precipitation validation values for each year $T$ are also shown  (using longer horizontal line segments).

In most cases, the validation values are within the distributions generated by KCDE-ITS simulation.   For the wet-season total precipitation, the simulations generated reliable predictions at all locations with the exception of the year 2019 ($T=41$).  In 2019 the data at the first three stations, which are located on the Western side of the Crete, exceed the range generated by KCDE-ITS.  The year 2019 was marked by  extremely heavy precipitation on Crete~\cite{HNMS}, and the western part of the island typically receives most of the precipitation.  Hence, it is not surprising that the simulation, which was trained on previous years of less intense precipitation, did not adequately capture the behavior for 2019.

Regarding estimates of the expected daily maximum precipitation,  the 95\% percentile of the ITS simulations seems adequate. For the six studied locations and the five validation years, the validation ERA5 maximum value exceeded the 95\% interval only in Chania ($s=1$) during 2019 ($T=41$).


Note that precipitation estimates at different locations are spatially correlated. Thus, 2016 was a rather dry year while 2019 a particularly wet year across the island.  Furthermore, the failure to adequately capture the rainfall amounts during 2019 at some of the test locations demonstrates a limitation of non-parametric methods---the extrapolation outside the range of the dataset can be  poor. According to~\cite{Banfi21}, in order to  reduce extrapolation errors samples longer than 50 years should be used.

\section{Discussion and Conclusions}
\label{sec:conclusions}

We propose the  KCDE approach for estimating the CDF of skewed probability distributions. Such distributions typically describe the  amount of  precipitation received by a geographical area over a given interval.  KCDE uses a non-parametric estimate of the CDF which is obtained by means of explicit integration of underlying kernel functions.
We showed that KCDE equipped with the BGK plug-in bandwidth~\cite{Botev10} estimates  skewed probability distributions  more accurately  than  the empirical staircase estimator or  KDE with the normal reference rule (Section~\ref{sec:synth_data} and~\ref{appC}). These results suggest that the BGK bandwidth, which is based on rigorous statistical theory, can efficiently adapt to the shape of different data probability distributions.
To our knowledge this is the first application of the BGK bandwidth in the non-parametric analysis of precipitation. As shown in Section~\ref{ssec:KDCE_Comparison_Staircase} the flexible BGK bandwidth in combination with compactly supported kernels in several cases out-performs the infinitely extended Gaussian kernel in the CDF estimation  of skewed probability distributions.

A different non-parametric estimator of the distribution of precipitation amount was used in~\cite{Mosthaf17}.  This work employed the Gaussian kernel with simpler plug-in bandwidths. The latter in most cases do not perform as well as  the BGK bandwidth according to~\cite{Botev10} and our analysis in Section~\ref{sec:synth_data}.  The non-parametric approach in~\cite{Mosthaf17} also uses CDF estimates, which are obtained by integrating  numerically the kernel density estimates.  In our approach, we explicitly integrate the kernel functions and provide analytical  expressions for the CDF steps for different kernels, both compactly supported and infinitely extended.

Our analysis and validation of  KCDE employed  synthetic data from parametric probability distributions and ERA5 reanalysis data.  We did not include observation data in this study because the available record for Crete is very sparse, with a large fraction of missing values~\cite{Agou2019}. In  the reanalysis data sets  we focused on the wet season for Crete.
Hence, we did not attempt to model the occurrence of precipitation. This can be achieved with Markov chain methods~\cite{Harrold03b}, or by means of censored latent Gaussian processes~\cite{Baxevani,Allard15}. However, it is not necessary if the goal is to model the precipitation amount over a certain period.

Taking advantage of the explicit equations for the CDF kernel steps, we formulated an ITS simulation algorithm based on KCDE which is suitable for non-Gaussian distributions. KCDE-ITS is a flexible and easy to use simulation method which can adapt to different distribution shapes.
In its current form, KCDE-ITS  does not account for correlations between precipitation amounts at different times.  Precipitation data, however, exhibit short-range correlations  (e.g., over a couple of days for the daily amounts) as evidenced in the variogram function (not shown herein). Such dependence  can be introduced in the simulations \emph{a posteriori}: A Monte Carlo approach can be used in which simulated values are shuffled aiming to optimize a correlation cost functional. This procedure is computationally expensive and also unnecessary for  estimating  total precipitation amounts over a given period of time.
For  methods that introduce space-time correlations in precipitation simulations we refer readers to~\cite{Papalexiou21}.
Regarding the use of KCDE-ITS simulations to design  protective measures against extreme precipitation, our investigations show the  95\% percentile of maximum daily precipitation predicted by KCDE-ITS based on 1000 simulated states provides an adequate estimate for most years.

It is generally accepted that in non-parametric estimation  no single kernel function or bandwidth estimate is optimal for all purposes.  We have found that the BGK bandwidth has advantages for the probability models studied. However, this finding cannot be generalized to all possible models of  distribution functions without further studies to determine optimality conditions for the BGK bandwidth. Finally, to successfully apply KCDE for simulation purposes, training data over a long  observation window are needed in order to capture rare events.  This is exemplified by the unsatisfactory performance of the KCDE-ITS simulation for the extreme-precipitation year 2019 as discussed in Section~\ref{sssec:era5-simulations}. The  issue of adequate sample size becomes even more pertinent  in light of climate change which is expected to significantly influence precipitation patterns in the Mediterranean basin.

The proposed KCDE method for non-parametric estimation of  probability distributions and the associated KCDE-ITS simulation method can be applied to  precipitation amounts in different geographical areas and climates. The reanalysis data used in this study focused on precipitation modeling for the Mediterranean island of Crete.  However, the
synthetic data were based on probability distributions that are commonly used in studies of precipitation.  The KCDE method can also be applied to other non-Gaussian hydrological variables (e.g., annual mean river flows, flood frequency, and droughts). An open goal (common to all non-parametric methods) is the enhancement of KCDE with improved extrapolation capability outside the range of  the observation dataset.

\section*{Acknowledgments}
This research is co-financed by Greece and the European Union (European Social Fund- ESF) through the Operational Programme \emph{Human Resources Development, Education and Lifelong Learning 2014-2020} in the context of the project ``Gaussian Anamorphosis with Kernel Estimators for Spatially Distributed Data and Time Series and Applications in the Analysis of Precipitation'' (MIS 5052133).




 \appendix

\clearpage
\newpage
\section{Integrals for CDF kernel steps}
\label{appA}
The change of variables $(z-z_{[i]})/h \to u_{i}$  implies $dz=  h\, du_{i}.$
Then, the CDF kernel step defined by Eq.~\eqref{eq:Kernel_integral} is expressed as
\begin{equation}
\tilde{K}\left(u_{i} \right)  = \int_{-\infty}^{u_{i}} du' K\left(u' \right) \,.
\label{Eq:Ap_Kernel_integral_u}
\end{equation}

Compact kernels are of the general form
$K(u)= g(u)\, \ind(|u|\le 1)$, where  $\ind(|u|\le 1)$ is the indicator function: $\ind(|u|\le 1)=1$ if $ \vert u \vert\leq1 $ and $\ind(|u|\le 1)=0$ otherwise.
For compact kernels, Eq.~\eqref{Eq:Ap_Kernel_integral_u} is expressed as follows
\begin{equation}
\tilde{K}\left(u_{i} \right)  =  \ind(u_{i}>- 1)\, \int_{-1}^{\ui} du' g\left(u' \right) \,,
\label{Eq:Ap_CompactKernel_integral_u}
\end{equation}
where  $\ui=\min(1,u_{i})$.
The above implies that the CDF kernel step vanishes if $u_{i}<-1$ for compact kernels. This reflects the fact that  the sample value $z_{[i]}$ does not contribute to the CDF at $z$, since the latter is more than one bandwidth less than $z_{[i]}$.
Below we explicitly calculate the integrals~\eqref{Eq:Ap_CompactKernel_integral_u} for the Epanechnikov and Bitriangular kernels.  The CDF steps for other kernels are similarly calculated (see also Table~\ref{tab:Kernels}).


\subsection{Epanechnikov kernel}


The Epanechnikov kernel is defined by means of the equation
\begin{equation}
K(u)=\frac{3}{4}(1-u^2)\, \ind(\vert u \vert \leq 1).
 \end{equation}

 Thus, the CDF kernel step for $u_{i}>-1$ is given by
 \begin{align}
\tilde{K}(u_{i})= &  \frac{3}{4}\int_{-1}^{\ui} du\, (1-u)^2 = \frac{3}{4} \left[ u - \frac{u^3}{3} \right]_{-1}^{\ui}
\nonumber \\
 = & \frac{3}{4} \left( \ui - \frac{1}{3} \ui^3 +\frac{2}{3} \right).
 \end{align}

\subsection{Bitriangular kernel}

The Bitriangular kernel is defined by means of the equation,

 \begin{equation}
     K(u)=\dfrac{3}{2}(1-\vert u \vert)^2 \, \ind_{\vert u \vert \leq 1} (u).
 \end{equation}

Thus, the CDF kernel step for $u_{i}>-1$ is given by
 \begin{align}
\tilde{K}(u_{i})= &  \frac{3}{2}\int_{-1}^{\ui} du\, (1-\vert u \vert)^2 =
\frac{3}{2} \left[
\dfrac{u_i^3}{3}-  u_i\lvert u_i\rvert + u_i
\right]_{-1}^{\ui}
\nonumber \\
 = & \frac{3}{2} \left[ \,
  \dfrac{\ui^3}{3} - \ui^2\,\sgn(u_i)
       + \ui +\dfrac{1}{3} \,
 \right].
 \end{align}


\clearpage
\section{BGK Bandwidth Estimation}
\label{appB}
This appendix reviews the details of BGK bandwidth estimation.
The estimate employs an auxiliary bandwidth $\hat{a}_{1}$, which is obtained from a sequence of bandwidths $a_{j;n}$, $j = l, l-1, \ldots, 1$. The index $n=0, 1, \ldots, n^{\ast}$ counts the number of repetitions until convergence is achieved, while  $l$ is a pre-selected  recursion depth (usually $l=7$).

The sequence is initialized by arbitrarily setting $a_{l+1;0}$ equal to the machine precision; the $a_{j;n}$ for given $n$ are generated by means of the following recursive expression
\begin{equation}
    \label{eq:botev_a_j}
    a^2_{j;n} = \left(
    \dfrac{1 + 2^{-(j+1/2)} }{3} \,
    \dfrac{1\times 3 \dots \times (2j-1)}{N \sqrt{\pi /2}\, G[\hat{f}_n^{(j+1)}(z)] }
    \right)^{ 2/(2j+3)},  \; j=l, l-1, \ldots, 1.
\end{equation}

\noindent In the above, $G[\hat{f}^{(j+1)}_{n}(\cdot)]$ is defined in Eq.~\eqref{eq:h_opt},  and $\hat{f}_n^{(j+1)}(\cdot)$ is the kernel-based PDF estimate with bandwidth $a_{j+1;n}$; the latter requires recursive calculation of $\hat{f}_{n}^{(j+2)}(\cdot)$.
If  $a_{j=l;n}$ is known, then $G[\hat{f}_n^{(l)}(z)]$ is also known, and  the $a_{j;n}, l  > j \ge 1$ follow by iterative  application of  Eq.~\eqref{eq:botev_a_j} for progressively smaller $j$.

For $n=0$ it is assumed that the initial bandwidth  $a_{l;0}$ is estimated using as  $\hat{f}^{(l+1)}_{n}(\cdot)$ in Eq.~\eqref{eq:botev_a_j} a Gaussian PDF with mean and variance estimated from the data.
One can then estimate the auxiliary bandwidth from Eq.~\eqref{eq:botev_a_j} for $n=0$ without further repetition.  However, as shown in \cite{Botev10}, if the true distribution deviates significantly from the normal, suboptimal estimates of $h$ are  obtained  from this initial condition. Better results are obtained by progressively refining the auxiliary bandwidth estimates.

For $n\ge 1$  Eq.~\eqref{eq:botev_a_j} is initialized using as  PDF  (for $j=l$) the kernel density estimate with bandwidth $a_{1;n-1}$, i.e., the PDF  $\hat{f}^{(1)}_{n-1}(\cdot)$ from the previous stage. The repetition counter  increases up to  $n^{\ast}$ such that  $\lvert a_{1;n^{\ast}} - a_{1;n^{\ast}-1}\rvert < \epsilon$.
Finally,  by setting $\hat{a}_{1}= a_{1;n^{\ast}}$  the  BGK bandwidth $h$ is calculated from Eq.~\eqref{eq:h_opt} and $\hat{a}_{1}$.




\section{KCDE  with Normal Reference  Bandwidth}
\label{appC}
\renewcommand{\thefigure}{C\arabic{figure}}
\setcounter{figure}{0}

\begin{figure}[h!]
\centering
    \begin{subfigure}[h!]{0.49\textwidth}
        \includegraphics[width=\textwidth]{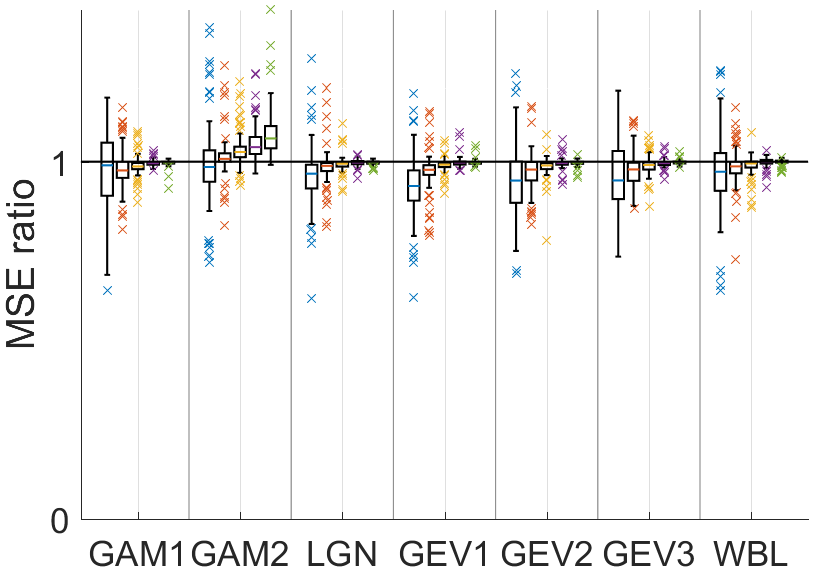}
        \vspace{-20pt}
        \caption{ Bitriangular \label{fig:RE_Epan_RoT} }
    \end{subfigure}
    \begin{subfigure}[h!]{0.49\textwidth}
        \includegraphics[width=\textwidth]{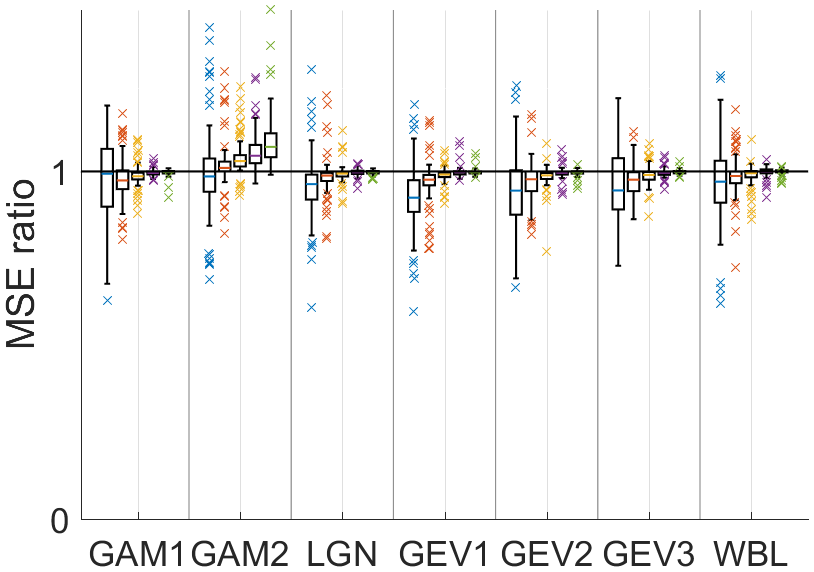}
        \vspace{-20pt}
        \caption{ Triweight \label{fig:RE_Triw_RoT} }
    \end{subfigure}
    \begin{subfigure}[h!]{0.49\textwidth}
        \includegraphics[width=\textwidth]{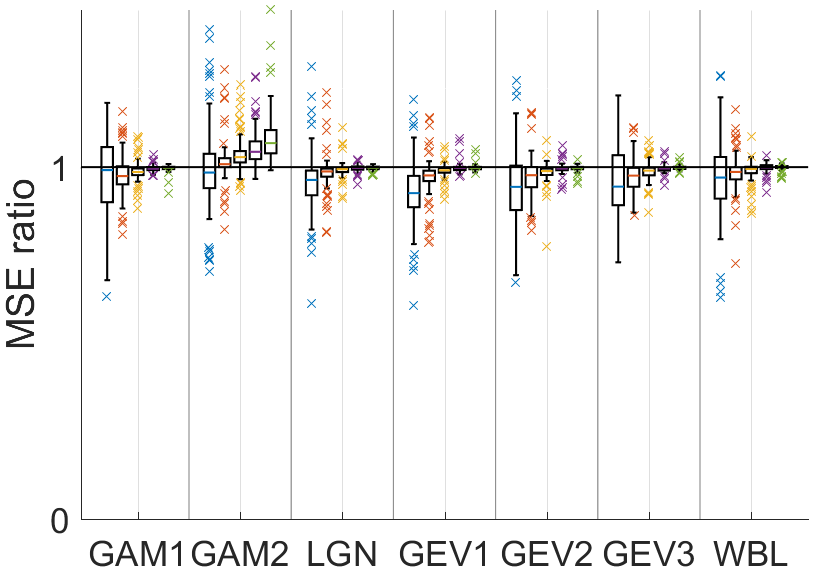}
        \vspace{-20pt}
        \label{fig:RE_Sphe_RoT}
        \caption{ Spherical   }
    \end{subfigure}
    \begin{subfigure}[h!]{0.49\textwidth}
        \includegraphics[width=\textwidth]{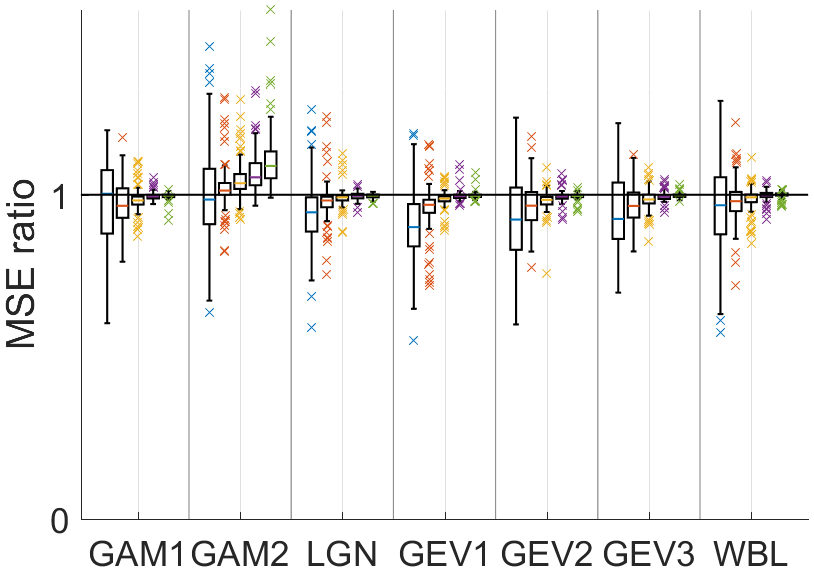}
        \vspace{-20pt}
        \label{fig:RE_Quad_RoT}
        \caption{ Epanechnikov }
    \end{subfigure}
    \begin{subfigure}[h!]{0.49\textwidth}
        \includegraphics[width=\textwidth]{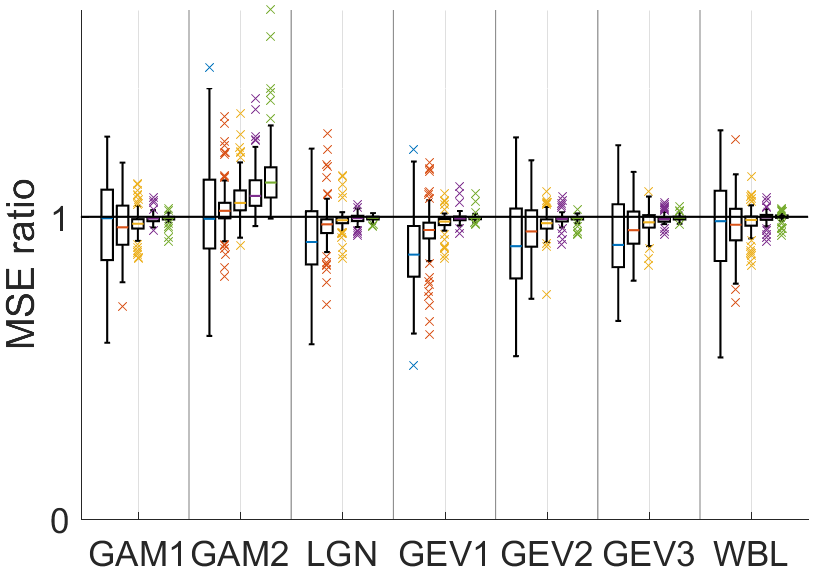}
        \vspace{-20pt}
        \label{fig:RE_Unif_RoT}
        \caption{ Uniform    }
    \end{subfigure}
    \begin{subfigure}[h!]{0.49\textwidth}

        \includegraphics[width=\textwidth]{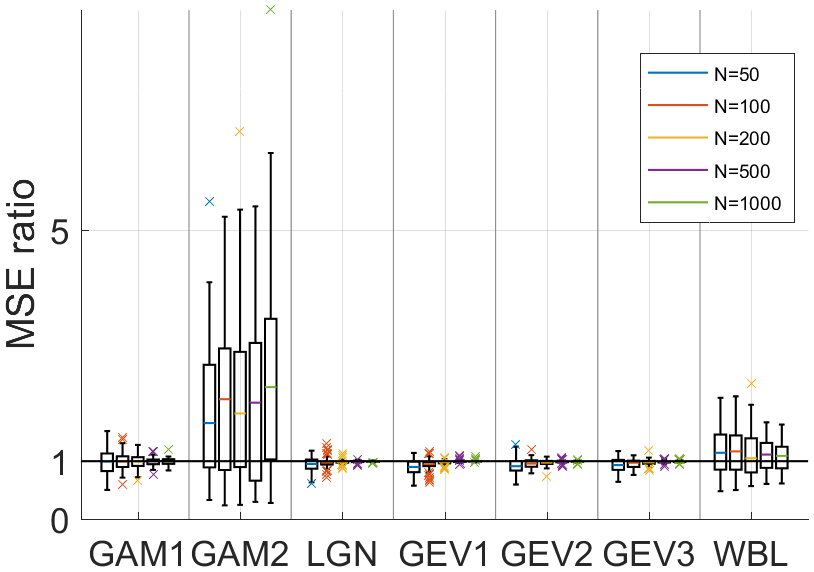}
        \vspace{-20pt}
        \label{fig:RE_Gaus_RoT}
        \caption{ Gaussian }
    \end{subfigure}
\caption{Boxplots of   RE\textsubscript{k|s} (MSE of KCDE $\Fk$ divided by the MSE of the staircase  estimate $\hat{F}_s$) using the Normal Reference Rule as bandwidth instead of the BGK. $N$ is the sample size. Abbreviations  are defined in the main text.}
\label{fig:RoT_Kern}
 \end{figure}

This appendix presents boxplots of the mean square error ratios  that compare the performance of KCDE  using the normal reference rule for bandwidth selection (see Section~\ref{ssec:CDF_compar}). The plots in Fig.~\ref{fig:RoT_Kern} illustrate that the CDF estimated with the normal reference rule does not significantly improve  the estimation compared to the staircase function.

\clearpage

\section{Exploratory Analysis of ERA5 Precipitation} 
\label{appD}
\renewcommand{\thetable}{D\arabic{table}}
\setcounter{table}{0}

\begin{table}[!ht]
\centering
\caption{Best-fit probability distribution models  for the daily, weekly, monthly and annual precipitation amounts for the wet period (October to March) (from ERA5 reanalysis data)  for the selected node near Messara valley. Measures of fit:  Akaike's Information Criterion (AIC), the Bayesian Information Criterion (BIC), and the negative log-likelihood (NLL). Measures of fit for the optimal models are boldfaced.
}
\label{table:Goddnessoffit_MonthlyReanalysis_N34_wet_nonzeros}
\resizebox{1\textwidth}{!}{%
\renewcommand\tabcolsep{3pt} 
\renewcommand\arraystretch{1} 
\begin{tabular}{ l| c c c c c c c c c}
\toprule
\diagbox{ {Method}}{ {Model}} &  {Normal} & {Weibull} & {Gamma} &  {GEV} & {LGN} \\
\midrule
\multicolumn{6}{c}{\textbf{Daily precipitation}}     \\
\midrule
\midrule
AIC & 3.2873$\,\times 10^4$	& \textbf{1.0878$\,\times 10^4$} &	1.1104$\,\times 10^4$ &	1.2876$\,\times 10^4$ &	1.1434$\,\times 10^4$\\
BIC & 3.2886$\,\times 10^4$  & \textbf{1.0892$\,\times 10^4$} & 1.1117$\,\times 10^4$ & 1.2896$\,\times 10^4$   & 1.1447$\,\times 10^4$ \\
NLL & 1.6434$\,\times 10^4$ & \textbf{5.4372$\,\times 10^3$} & 5.5501$\,\times 10^3$ & 6.4352$\,\times 10^3$ & 5.7149$\,\times 10^3$ \\
\midrule
\multicolumn{6}{c}{\textbf{Weekly precipitation}}     \\ %
\midrule\midrule
AIC    & 8.3461$\,\times 10^3$  & 6.9989$\,\times 10^3$  & \textbf{6.9412$\,\times 10^3$}  & 7.4841$\,\times 10^3$  & 7.4278$\,\times 10^3$  \\
BIC    & 8.3560$\,\times 10^3$  & 7.0089$\,\times 10^3$  & \textbf{6.9511$\,\times 10^3$}  & 7.4990$\,\times 10^3$  & 7.4377$\,\times 10^3$  \\
NLL    & 4.1710$\,\times 10^3$  & 3.4975$\,\times 10^3$  & \textbf{3.4686$\,\times 10^3$}  & 3.7391$\,\times 10^3$  & 3.7119$\,\times 10^3$  \\
\midrule
\multicolumn{6}{c}{\textbf{Monthly precipitation}}     \\
\midrule\midrule
AIC & 2.3649$\,\times 10^3$  & \textbf{2.3276$\,\times 10^3$}  & 2.3428$\,\times 10^3$  & 2.3406$\,\times 10^3$  & 2.3996$\,\times 10^3$ \\
BIC & 2.3719$\,\times 10^3$  & \textbf{2.3346$\,\times 10^3$}  & 2.3467$\,\times 10^3$  & 2.3531$\,\times 10^3$  & 2.4066$\,\times 10^3$ \\
NLL & 1.1832$\,\times 10^3$  & \textbf{1.1618$\,\times 10^3$}  & 1.1678$\,\times 10^3$  & 1.1673$\,\times 10^3$  & 1.1988$\,\times 10^3$ \\
\midrule
\multicolumn{6}{c}{\textbf{Annual precipitation}}     \\
\midrule\midrule
AIC & 463.8558  & \textbf{463.4259} & 465.4719 & 467.1305   &  465.8434 \\
BIC & 467.2829  & \textbf{466.8531} & 468.8991 & 470.8069   & 470.6502 \\
NLL & 230.7360 & \textbf{229.7130} & 231.5652 & 229.9199 & 229.9217 \\
\bottomrule
\end{tabular}
}
\end{table}

\begin{table}[!ht]
\centering
\caption{Number of times $\in \{0, 1, 2 \ldots, 65\}$ each  parametric model is selected as the optimal distribution  for the daily, weekly, monthly and annual precipitation amounts for the wet period (October to March) for the 65  ERA5 nodes around Crete.    Measures of fit: Akaike's Information Criterion (AIC), the Bayesian Information Criterion (BIC), and the negative log-likelihood (NLL). Measures of fit for the optimal models are boldfaced.
}
\label{table:best_models_all_nodes_wet_nonzeros}
\begin{tabular}{ l| c c c c c c c c c}
\toprule
\diagbox{ {Method}}{ {Model}} & Normal & Weibull & Gamma &  GEV & LGN \\
\midrule
\multicolumn{6}{c}{Daily precipitation}     \\
\midrule
\midrule
AIC & 0  & {56} & 9 &  0   &  0 \\
BIC & 0  & {56} & 9 &  0   &  0 \\
NLL & 0  & {56} & 9 &  0   &  0 \\
\midrule
\multicolumn{6}{c}{Weekly precipitation}     \\
\midrule\midrule
AIC & 0  & 0 & {65} &  0   &  0 \\
BIC & 0  & 0 & {65} &  0   &  0 \\
NLL &  0  & 0 & {65} &  0   &  0 \\
\midrule
\multicolumn{6}{c}{Monthly precipitation}     \\
\midrule\midrule
AIC & 0  & {61} &  4 & 0   & 0 \\
BIC & 0  & {61} &  4 & 0   & 0 \\
NLL & 0  & {59} &  1 & 5   & 0 \\
\midrule
\multicolumn{6}{c}{Annual precipitation}     \\
\midrule\midrule
AIC &  18  & 5 & {28} & 0   &  14 \\
BIC &  18  & 5 & {28} & 0   &  14 \\
NLL & 3 & 2 & 7 &  {52} & 1 \\
\bottomrule
\end{tabular}
\end{table}

\medskip

\end{document}